\documentclass[conference]{IEEEtran}
\IEEEoverridecommandlockouts

\usepackage{cite}
\usepackage{amsmath,amssymb,amsfonts}
\usepackage{algorithmic}
\usepackage{graphicx}
\usepackage{textcomp}
\usepackage{booktabs}
\usepackage{tikz}
\usepackage{array}
\usepackage{tabularx, booktabs}
\usepackage{siunitx}
\usepackage{xcolor}
\usepackage{caption}
\usepackage{multirow}
\usepackage{relsize}
\usepackage{soul}\usepackage[colorinlistoftodos]{todonotes}
\usepackage{siunitx}
\def\BibTeX{{\rm B\kern-.05em{\sc i\kern-.025em b}\kern-.08em
    T\kern-.1667em\lower.7ex\hbox{E}\kern-.125emX}}
\usepackage{subcaption}
\usepackage{hyperref}
\usepackage{listings}    
\usepackage{xcolor}      
\usepackage{courier}     
\definecolor{codegray}{rgb}{0.95,0.95,0.95}
\definecolor{codeblue}{rgb}{0,0,0.8}

\newcolumntype{C}[1]{>{\centering\arraybackslash}m{#1}}

\newcommand{\configcell}[2]{%
  \begin{tabular}[c]{@{}c@{}}%
    #1 \\[3pt]%
    {\small\textit{#2}}%
  \end{tabular}%
}

\tikzset{atom/.style={circle, fill=blue!65!black, inner sep=0pt, minimum size=5pt}}

\begin{document}

\title{A Transferable Machine Learning Approach to Predict Optimized Orbitals for Electronic Structure Problems%
\thanks{The research is part of the Munich Quantum Valley (MQV), which is supported by the Bavarian state government with funds from the Hightech Agenda Bayern Plus.\\
Email address for correspondence:\\
abhishek.yogendra.dubey@iis.fraunhofer.de}}

\author{
    \IEEEauthorblockN{
        Lucas van der Horst\IEEEauthorrefmark{1}, 
        Maniraman Periyasamy\IEEEauthorrefmark{1}, 
        Abhishek Y. Dubey\IEEEauthorrefmark{1}, \\
        Davide Bincoletto\IEEEauthorrefmark{2}, 
        Jakob S. Kottmann\IEEEauthorrefmark{2}, and 
        Daniel D. Scherer\IEEEauthorrefmark{1}
    }
    \IEEEauthorblockA{
        \IEEEauthorrefmark{1}\textit{Fraunhofer Institute for Integrated Circuits IIS}, Nürnberg, Germany
    }
    \IEEEauthorblockA{
        \IEEEauthorrefmark{2}\textit{Institute for Computer Science}, \textit{University of Augsburg}, Augsburg, Germany
    }
}

\maketitle

\begin{abstract}
Variational quantum eigensolver ansätze hold considerable promise for ground-state energy calculations on near-term quantum hardware, yet most promising ansatz designs currently strongly depend on how well the molecular orbital basis captures the electronic correlation of the system. Computing optimized orbital coefficients via classical routines is computationally expensive and must be performed independently for each molecular geometry -- a bottleneck that limits scalability across chemical space. We present a graph neural network framework that predicts optimized orbital coefficients directly from molecular geometry and pair-wise bonding structure. Trained on hydrogenic systems of modest size ($H_4$ and $H_6$) across tens of thousands of geometries, our model transfers to larger, unseen systems ($H_8$, $H_{10}$ and $H_{12}$) without retraining -- demonstrating strong out-of-distribution generalization with respect to system size. When evaluating on structured and random configurations, and comparing against energies obtained with full classical optimization, our model reaches mean absolute energy errors $\mathcal{O}(10^2)$ and $\mathcal{O}(10)$ milli-Hartrees, respectively. Beyond energy estimation, the predicted orbitals serve as high-quality warm-start initializations that substantially reduce optimizer iterations to ground-state energy convergence.
These results establish graph neural networks as an effective and scalable strategy for accelerating orbital optimization in hybrid quantum-classical workflows, directly reducing the classical pre-processing overhead that currently limits the practical deployment of variational quantum eigensolver on near-term quantum hardware.
\end{abstract}

\begin{IEEEkeywords}
VQE, GNN, Transferable Machine Learning, Orbital Optimization 
\end{IEEEkeywords}

\section{Introduction}
\label{sec:Intro}
Quantum simulation and quantum chemistry are widely  expected to be among the first fields to experience transformative impact from digital quantum computers~\cite{feynman2018simulating} with applications spanning pharmaceutical drug discovery~\cite{santagati2024drug} and the design of next-generation catalysts and battery materials~\cite{zini2023quantum, delgado2022simulating}. Central to these applications is the electronic structure problem: determining the ground and low-lying excited states of a system of interacting electrons. Classical methods have advanced considerably in addressing this problem~\cite{saad2010numerical}, yet they face a fundamental and unavoidable barrier--the exponential cost of representing a many-body wavefunction on a classical hardware. This bottleneck becomes especially acute for large, strongly correlated molecular systems, where classical approximations break down. Quantum computers offer a structurally different approach: a quantum processor encodes the many-body state natively in its qubit register, making simulation tractable in principle. Quantum phase estimation (QPE) algorithm~\cite{kitaev1995quantum, abrams1999quantum, aspuru2005simulated} is the primary proposal for estimating energy eigenvalues with $\mathcal{O}(n)$ qubits for an $n$ electron system. However, QPE faces the ``orthogonality catastrophe", where the overlap between a classically prepared initial state and the true ground state decays exponentially with system size~\cite{tubman2018postponing}, making high-quality state preparation a prerequisite for algorithmic tractability.

The Variational Quantum Eigensolver (VQE)~\cite{peruzzo2014variational} has emerged as a practical bridge between Noisy Intermediate-Scale Quantum (NISQ)~\cite{preskill2018quantum} and Fault-tolerant Quantum computing (FTQC). As a hybrid quantum-classical algorithm, VQE iteratively refines a parametrized trial state via the Rayleigh-Ritz variational principle~\cite{gross1988rayleigh}, yielding noise-resilient energy estimates on near-term hardware while producing high-overlap trial states suitable as initial input for future fault-tolerant QPE algorithms -- directly mitigating the orthogonality catastrophe. The accuracy of VQE depends critically on the ansatz. Physically motivated ans{\"a}tze such as Unitary Coupled Cluster (UCC) offer high accuracy~\cite{romero2019strategies, tilly2022variational, anand2022quantum} but frequently exceed the coherence limits of current hardware. The Separable Pair Approximation (SPA) ansatz ~\cite{kottmann2022optimized} addresses this trade-off by modeling the wavefunction as a product of independent electron pairs, achieving linear memory scaling and reduced circuit depth. To further improve energy estimation, Orbital Optimization (OO)~\cite{mizukami2020orbital} is incorporated into the SPA workflow. OO treats the spatial orbitals themselves as variational parameters, allowing the molecular orbital basis to relax and adapt to the electronic correlation. This enables higher accuracy by concentrating correlation into a compact orbital basis; it reduces gate-count by replacing single-excitation operators with equivalent orbital rotations; and it improves the overlap with the true ground state, directly mitigating the orthogonality catastrophe in future fault-tolerant settings~\cite{mizukami2020orbital, sokolov2020quantum, yalouz2021state}.

Despite these advantages, OO introduces a significant computational bottleneck that becomes increasingly severe as the quantum component of the pipeline scales up. Within SPA, OO remains tractable because the wavefunction admits efficient classical simulation. However, for a general quantum VQE ansatz, evaluating the energy gradient with respect to orbital rotation parameters requires repeated state overlap computations and nested optimization loops, both of which scale poorly with system size~\cite{mizukami2020orbital}. Crucially, this procedure must be repeated independently for every molecular geometry, with no mechanism for knowledge reuse across related systems.

We address this directly by proposing a graph neural network (GNN) based regression framework that learns to predict optimized orbital rotation parameters from molecular geometry and pairwise bonding structure alone, bypassing explicit classical optimization at inference time. The main contributions of this work are:
\begin{itemize}
    \item \textbf{Transferable orbital prediction:} A GNN-based regression model trained on small hydrogenic systems ($H_4$ and $H_6$) that generalizes to larger, unseen systems ($H_8, H_{10}$ and $H_{12}$) without retraining, demonstrating strong out-of-distribution generalization with respect to system size.
    \item \textbf{Dual-scale molecular embedding:} A dual-scale GNN architecture that processes molecular graphs at two complementary length scales to produce size-agnostic orbital representations.
    \item \textbf{Gauge-invariant training objective:} A physics-informed loss function combining a Huber regression term on the orbital generator matrix with gauge-invariant subspace and orbital overlap losses, ensuring robustness to the sign and permutation ambiguities intrinsic to molecular orbital representations.
    \item \textbf{Warm-start initialization:} Even where direct orbital prediction carries moderate error, the predicted matrices serve as high-quality warm-start initializations that substantially reduce classical optimizer iterations to convergence, extending the practical utility of the model beyond the regime of direct prediction.
\end{itemize}

Critically, by replacing the most expensive classical component of the VQE pipeline with a fast learned surrogate, this work directly addresses a practical bottleneck in near-term quantum computing: the classical co-processor overhead that currently prevents VQE from scaling to industrially relevant molecular systems on NISQ devices.

The remainder of this paper is organized as follows: Section~\ref{sec:PrevWork} reviews the relevant previous work on machine learning methods used for chemical problems. Section~\ref{sec:Methodology} describes the data generation process for the supervised learning task, the engineering of the features and model, and finally the training method including the loss functions. Section~\ref{sec:Eval} reports evaluation results using our model on structured and random geometric configuration on unseen dataset demonstrating transferability. Finally, Section~\ref{sec:Conc} concludes with the limitations of the current approach and directions for future work.

\begin{table}[ht] 
  \centering
  \setlength{\tabcolsep}{2pt} 
  \renewcommand{\arraystretch}{1.4} 

  \caption{%
    Geometric configurations and sizes of the training dataset.
    Filled circles represent atomic positions.
  }
  \label{tab:dataset_configs}

  \begin{tabularx}{\columnwidth}{|>{\centering\arraybackslash}p{1.4cm}| >{\centering\arraybackslash}X | >{\centering\arraybackslash}p{1.2cm}|}
    \toprule
    \textbf{Geom.}
      & \textbf{Configuration}
      & \textbf{Size} \\
    \midrule

    \multirow{2}{*}[-4pt]{\textbf{Linear}}
      & \configcell{%
          \begin{tikzpicture}[scale=0.60, baseline=-0.5ex]
            \draw[gray!65, densely dashed, thin] (-0.18,0) -- (1.68,0);
            \foreach \x in {0, 0.50, 1.00, 1.50}
              \node[atom] at (\x, 0) {};
          \end{tikzpicture}%
        }{Equidistant}
      & 50k \\[4pt]

      & \configcell{%
          \begin{tikzpicture}[scale=0.60, baseline=-0.5ex]
            \draw[gray!65, densely dashed, thin] (-0.18,0) -- (1.68,0);
            \foreach \x in {0, 0.33, 0.9, 1.50}
              \node[atom] at (\x, 0) {};
          \end{tikzpicture}%
        }{Random}
      & 10k \\

    \midrule

    \multirow{3}{*}[-20pt]{\textbf{Planar}}
      & \configcell{%
          \begin{tikzpicture}[scale=0.60, baseline=-0.5ex]
            \foreach \x in {0.10, 0.85}
              \foreach \y in {0.00, 0.70}
                \node[atom] at (\x, \y) {};
          \end{tikzpicture}%
        }{Equidistant}
      & 50k \\[4pt]

      & \configcell{%
          \begin{tikzpicture}[scale=0.60, baseline=-0.5ex]
            \node[atom] at (0.08, 0.62) {};
            \node[atom] at (0.62, 0.01) {};
            \node[atom] at (1.20, 0.72) {};
            \node[atom] at (0.42, 0.35) {};
          \end{tikzpicture}%
        }{Random}
      & 10k \\[4pt]

      & \configcell{%
          \begin{tikzpicture}[scale=0.60, baseline=-0.5ex]
            \draw[gray!65, densely dashed, thin] (0.55,0.55) circle (0.55);
            \foreach \a in {45, 135, 225, 315}
              \node[atom] at ({0.55 + 0.55*cos(\a)}, {0.55 + 0.55*sin(\a)}) {};
          \end{tikzpicture}%
        }{Ring}
      & 10k \\

    \midrule

    \textbf{3D}
      & \configcell{%
          \begin{tikzpicture}[scale=0.60, baseline=-0.5ex]
            \draw[gray!45, ->, thin] (0.50, 0.38) -- (1.18, 0.38);
            \draw[gray!45, ->, thin] (0.50, 0.38) -- (0.50, 1.05);
            \draw[gray!45, ->, thin] (0.50, 0.38) -- (0.10, 0.02);
            \node[circle, fill=blue!35!black!60, inner sep=0pt, minimum size=4pt] at (0.20, 0.72) {};
            \node[circle, fill=blue!35!black!60, inner sep=0pt, minimum size=4pt] at (0.32, 0.18) {};
            \node[atom] at (0.85, 0.62) {};
            \node[atom] at (1.1, 0.92) {};
          \end{tikzpicture}%
        }{3D Random}
      & 10k \\

    \bottomrule
    \multicolumn{2}{|c|}{\small\textit{Total}} & 140k \\
    \bottomrule
  \end{tabularx}
\end{table}

\section{Previous Work}
\label{sec:PrevWork}
\begin{figure*}[p]
\centering
\includegraphics[width=0.90\textwidth]{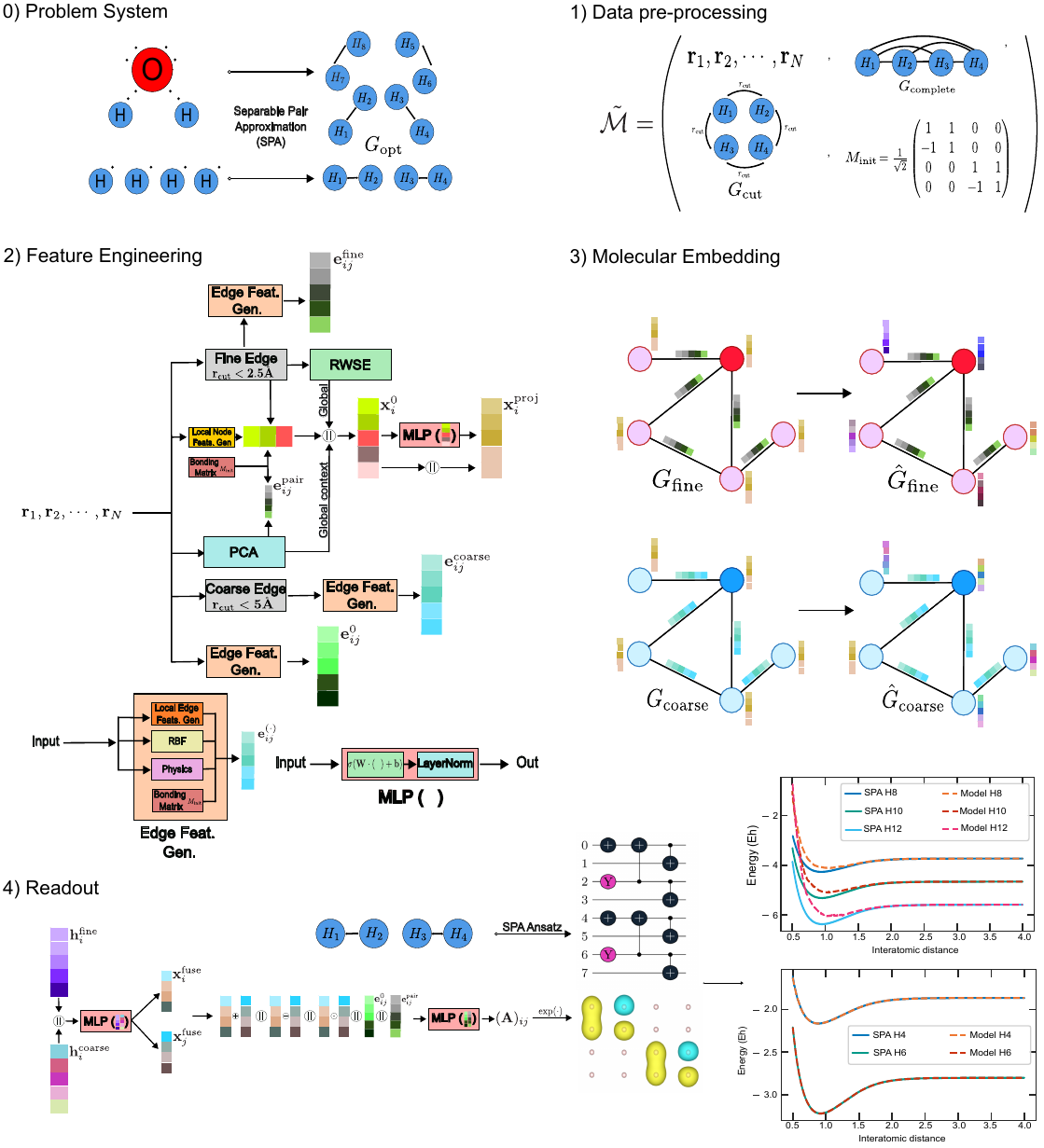}
\caption{Complete pipeline for transferable orbital prediction. \textbf{Stage 0)}: a molecular system is mapped to paired hydrogenic configurations under the Separable Pair Approximation (SPA), yielding a pairwise bonding graph $G_{\text{opt}}$, whose geometry is inherited directly from the physical atomic coordinates. \textbf{Stage 1)} Construct the enriched molecular representation $\tilde{\mathcal{M}}$ from three complementary graph structures: the complete graph $G_{\text{complete}}$, the radius graph $G_{\text{cut}}$ retaining only neighbors within the cutoff $r_{\text{cut}}$ and the Givens rotation guess matrix $M_{\text{init}}$ encoding the pairwise bonding structure. \textbf{Stage 2)} Construct physically motivated node and edge descriptors from the 3D atomic positions $\{\mathbf{r}_1, \ldots, \mathbf{r}_N\}$ and $M_{\text{init}}$, operating at two complimentary length scales -- fine ($r_{\text{fine}}=2.5$~\AA) and coarse ($r_{\text{coarse}}=5$~\AA) -- to capture both bond-level and medium-range structural context. \textbf{Stage 3)} Process these representations through two parallel GNNs, producing fine and coarse latent node embeddings $\hat{\mathcal{G}}_{\text{fine}}$ and $\hat{\mathcal{G}}_{\text{coarse}}$ which are subsequently fused into a unified molecular representation. \textbf{Stage 4)} Combine the fused node embeddings, pairwise with edge and geometric features and pass through a shared readout MLP, predicting the upper-triangular entries of the antisymmetric generator matrix $\textbf{A}_{\text{upper}}$. The full orbital rotation matrix is recovered as $M_{oo}=e^{\mathbf{A}}$, which is used to construct the SPA ansatz circuit and compute the ground-state energy of the system.}
\label{fig: Model Architecture}
\end{figure*}

Machine learning has emerged as a powerful complement to quantum chemistry, offering data-driven alternatives to computationally expensive electronic structure routines. Among the various architectural families explored for molecular property prediction, GNNs have proven particularly well-suited to the task: they operate directly on molecular graph representations, avoiding hand-crafted descriptors while naturally capturing inter-atomic interactions. Early work established foundational graph convolutional models, while subsequent contributions introduced geometry-aware architectures that are especially relevant for quantum chemistry tasks \cite{reiser2022graph}. Duvenaud et al. \cite{duvenaud2015convolutional} introduced learned molecular fingerprints via iterative neighborhood aggregation, and Gilmer et al. \cite{gilmer2017neural} formalized the Message Passing Neural Network (MPNN) framework, incorporating multidimensional edge features to encode chemical bonds and interatomic distances. A pivotal step towards quantum chemistry applicability came with SchNet~\cite{schutt2017schnet} which introduced continuous-filter convolutional network that operates on atomic numbers and 3D coordinates, yielding rotationally and translationally invariant molecular embeddings. Subsequent architectures -- including PhysNet \cite{unke2019physnet}, DimeNet \cite{gasteigerdirectional}, and GemNet \cite{gasteiger2021gemnet} -- progressively enriched geometric expressivity by incorporating angular and torsional information, improving both accuracy and physical interpretability.

A known limitation of standard MPNNs is their bounded expressivity: they are at most as powerful as the 1-Weisfeiler-Leman (1-WL) graph isomorphism test \cite{weisfeiler1968reduction}, meaning that they cannot distinguish non-isomorphic graphs under a single-hop neighborhood aggregation. This presents practical challenges for symmetric molecular configurations, such as equidistant linear or planar geometries, where distinct physical arrangements may produce identical graph representations. Two complementary strategies have been proposed to address this: enhancing expressivity through richer directional encodings~\cite{mao2025molecule}, and augmenting local message passing with positional and structural encodings alongside global attention mechanisms~\cite{rampavsek2022recipe}. The transferability of GNN-based models has also been demonstrated beyond molecular systems — notably in Rydberg atom arrays~\cite{simard2025learning}, where models trained on small configurations generalize to larger lattice sizes and unseen geometries, a result that motivates the size-transferability objective central to our work.

A complementary contribution directly relevant to the present work is that of Bincoletto et al.~\cite{bincoletto2025transferable}, who propose a transferable machine learning framework for predicting variational circuit parameters within the SPA ansatz. Their model maps molecular geometry and graph structure to SPA circuit parameters using Graph Attention Networks and SchNet-based embeddings, achieving accurate zero-shot generalization to larger hydrogen chain systems up to $H_{12}$. While their approach targets the variational parameters governing the quantum circuit directly, our method addresses the upstream component of the same workflow: predicting the orbital coefficient matrix used to rotate the molecular orbital basis prior to circuit construction. Together, these two approaches form a unified data-driven pipeline — one replacing the orbital optimization step, the other replacing variational parameter search — with the shared goal of reducing the classical computational overhead in hybrid quantum-classical algorithms.

\section{Methodology}
\label{sec:Methodology}
This section describes the end-to-end pipeline for training the proposed GNN-based orbital prediction framework. We first detail the data generation procedure used to construct the training corpus, followed by the target representation and pre-processing steps. We then present the model architecture and conclude with the training objective.

\subsection{Data generation}
 
Training data is generated using the \textsc{quanti-gin} \cite{steinNylser2025} library which uses the \textsc{Tequila} framework~\cite{kottmann2021tequila} -- an open-source quantum chemistry module designed for rapid prototyping of variational quantum algorithms. We construct a dataset of hydrogen configurations $H_N$ $\left(N \in \{4,6,8,10,12\}\right)$ by running the full orbital-optimized VQE pipeline across a large and geometrically diverse set of molecular configurations. Each data point is produced through a four-stage procedure: geometry sampling, chemical graph selection, orbital-optimization and SPA energy minimization.

To promote geometric diversity and support transferability, atomic coordinates are sampled from several geometric families subject to the constraint that the nearest neighbor spacing lies in the range $(0.5, 4.0)$ \AA, \ keeping all configurations near the ground-state regime. The geometry families and corresponding dataset sizes per molecule size are reported in Table~\ref{tab:dataset_configs}.

The data generation procedure follows~\cite{bincoletto2025transferable} and is enumerated as follows:

\begin{enumerate}
  \item Sample atomic coordinates $\mathbf{R}\in \mathbb{R}^{N \times 3} = \{\mathbf{r}_i\}_{i=1}^N$ from one of the geometry families from Table~\ref{tab:dataset_configs}.
  \item Determine the optimal chemical graph $G_\text{opt}$ heuristically as the perfect matching graph with minimal total edge weight using scaled Euclidean distance as edge weights.
  \item Construct a structured initial guess for the orbital coefficients from $G_\text{opt}$: for each edge $(i,j)$, the $2\times 2$ block indexed by $\{(i,i),(i,j),(j,i),(j,j)\}$ is set to the Givens rotation $M_{\text{init}}=\frac{1}{\sqrt{2}}\bigl[\begin{smallmatrix}1 & 1 \\ -1 & 1\end{smallmatrix}\bigr]$, mimicking a bonding / anti-bonding decomposition of the electron pair.
  \item Construct the SPA ansatz ($U_{\text{SPA}}$) for the graph $G_\text{opt}$ and perform VQE energy minimization, simultaneously optimizing the circuit parameters $\boldsymbol{\theta}$ (see Appendix A in~\cite{bincoletto2025transferable}) and the orbital matrix $M_{\text{oo}}$.
  \item  With the optimized orbitals fixed, assemble the molecular Hamiltonian ($\hat{H}$) and evaluate the SPA energy $E_{\text{SPA}}=\min_{\boldsymbol{\theta}}\langle U_{\text{SPA}}(\boldsymbol{\theta})|\hat{H}| U_{\text{SPA}}(\boldsymbol{\theta}) \rangle$ recording the optimal circuit parameters $\boldsymbol{\theta}_{\text{opt}}$.
  \item Store the tuple $(\mathbf{R}, G_\text{opt}, \boldsymbol{\theta}_{\text{opt}},  M_{\text{oo}}, E_{\text{SPA}})$. Since the molecule can be completely defined by its coordinates and pair-wise graph structure, we represent this mathematically as $\mathcal{M}=(\mathbf{R}, G_\text{opt})$.
\end{enumerate}

\subsection{Data Format and Machine-Learning Target}

The orbital coefficient matrix $M_{\text{oo}}$ poses a non-trivial regression challenge: since the SPA ansatz employs real orthonormal basis functions from the STO-3G set, the assembled molecular orbital coefficients are real-valued~\cite{kottmann2023molecular}, and the resulting matrix is orthogonal.  

To resolve this, we apply the matrix logarithm in a pre-processing step, mapping $M_{\text{oo}}$ to its generator $\mathbf{A} = \log M_{\text{oo}}$. For orthogonal matrices, the matrix logarithm in this case is real and antisymmetric, transforming a constrained matrix-valued target into an unconstrained one. The model therefore predicts the $N(N-1)/2$ strictly upper-triangular entries $\mathbf{A}_{\mathrm{upper}}$, which encode all independent degrees of freedom. At inference time, the full antisymmetric matrix is reconstructed as $\mathbf{A} = \mathbf{A}_{\mathrm{upper}} -
\mathbf{A}_{\mathrm{upper}}^\top$, and the orthogonal orbital rotation is recovered via matrix exponentiation, $M_{\text{oo}} = e^{\mathbf{A}}$.

\subsection{Model Architecture}
\label{subsec:model_arch}

We seek a parametric map 
\begin{equation}
    F_\mathbf{\Theta}: \mathcal{M} \longmapsto M_\text{oo} \in \mathbb{R}^{N \times N},
\end{equation}
which we decompose into four stages: 1) data pre-processing, 2) feature engineering, 3) dual-scale molecular embedding and 4) readout function. The architecture is designed around the key requirement of \emph{size transferability}: every feature and every learned operation must be \emph{locally defined}, depending only on atom's neighborhood rather than the total system size. This ensures that learned representations
transfer across molecule sizes.

The backbone of $F_\mathbf{\Theta}$ is a GNN which provides a natural framework for learning on molecular structures and has shown strong performance in chemical problems \cite{kang2025orbitall,van2023casnet}. A GNN operates on a graph $G = \left(\mathcal{V}, \mathcal{E}, \{\mathbf{x}_i\}_{i=1}^N, \{\mathbf{e}_{ij}\}_{(i, j) \in \mathcal{E}}\right)$, where $\mathcal{V}=\{v_i\}_{i=1}^N$ denotes the set of nodes, $\mathcal{E}$ the set of edges, $\mathbf{x}_i \in \mathbb{R}^{d_x}$ the node features, and $\mathbf{e}_{ij} \in \mathbb{R}^{d_e}$ the edge features. Through $K$ rounds of message passing, each node accumulates information from its $K$-nearest neighbors, producing per-node latent representations $\mathbf{h}_i \in \mathbb{R}^{d_{\text{embed}}}$. By construction, GNNs are equivariant to node permutations — symmetry that the orbital coefficient matrix itself must respect, making GNNs a natural architectural choice. Pairwise euclidean distances $d_{ij}=||\mathbf{r}_i-\mathbf{r}_j||_2$ serve as the primary geometric features, as orbital coefficients are invariant to rigid-body transformations of the molecular frame.

\subsubsection{Data pre-processing}
The optimal graph $G_\text{opt}$ encodes only bonded pairs and, if used alone, would restrict message passing to directly bonded atoms. We therefore work with two complimentary connectivity graphs. The \textbf{complete graph} $G_{\text{complete}}$ connects every pair by a unique edge and is used during the readout to evaluate predictions for all $(i < j)$ pairs. The \textbf{radius graph} $G_{\text{cut}}$ connects pairs $i$ and $j$ whenever $||\mathbf{r}_i - \mathbf{r}_j||<r_{\text{cut}}$. and is used exclusively with the GNN message. Additionally, the initial \textbf{bonding guess matrix} $M_{\text{init}}$ provides the actual pair-wise bonding information.

The enriched molecular representation is thus 
\begin{equation}
    \tilde{\mathcal{M}} = (\mathbf{R}, G_{\text{complete}}, G_{\text{cut}}, M_{\text{init}}).
\end{equation}

\subsubsection{Feature Engineering}
Given $\tilde{\mathcal{M}}$, we construct physically motivated local descriptors for both nodes and edges before message passing begins.

\textbf{Node feature enhancement}. Node features are computed \emph{directly from 3D positions} $\mathbf{R}$ and are defined locally with respect to $G_{\text{cut}}$. The base feature set includes the atomic number $Z$, the log-scaled coordination number of the nodes within the cutoff region $r_{\text{cut}}$, and distributional statistics of neighboring distances.

\begin{table}[!t]
\renewcommand{\arraystretch}{1.3} 
\caption{Node features using in both the GNN tracks}
\label{tab:node_features}
\centering
\begin{tabularx}{\columnwidth}{l X} 
\toprule
\textbf{Feature} & \textbf{Description} \\
\midrule
$z$ & Atomic number (1 for hydrogen chains) \\
$\log(1 + \deg_i)$ & Log-scaled coordination number (neighbours within $r_{\text{cut}}$) \\
$\bar{d}_i$ & Mean neighbour distance \\
$\sigma_{d,i}$ & Standard deviation of neighbour distances \\
$d_{\min}/d_{\max}$ & Distance asymmetry ratio (min/max neighbour dist) \\
$\|\bar{\mathbf{v}}_i\|$ & Directional asymmetry: magnitude of mean displacement vector to neighbours \\
$\frac{\bar{d}_i - \frac{d_{\min}+d_{\max}}{2}}{\sigma_{d,i}}$ & Distance skewness proxy \\
$\bar{d}_{\text{pair}}$ & Distance to paired partner according to $G$ \\
$\mathbf{R}_{\text{PCA}}$ & PCA of centroid centered positions \\
$d_{\text{center}}$ & Centroid centered positions \\
$p_i^{\text{RWPE}}$ & Random-walk positional encoding~\cite{dwivedi2021graph} \\
\bottomrule
\end{tabularx}
\end{table}

These are augmented with symmetry-breaking descriptors that are critical for distinguishing geometrically degenerate configurations: the directional asymmetry ($\|\bar{\mathbf{v}}_i\|$) differentiates end atoms from interior atoms in equidistant linear chains, since end atoms have all neighbors on one side while interior atoms do not; the centroid distance $d_{\text{centroid}}$ provides an analogous interior/end distinction for planar and ring geometries. The bond-partner distance $\|\mathbf{r}_i - \mathbf{r}_{p(i)}\|$ encodes pairing information from $G_{\text{opt}}$, and PCA (principal component analysis)-projected positions $\mathbf{R}_{\text{PCA}}\in\mathbb{R}^3$ together with random-walk positional encoding $p_i^{\text{RWPE}} \in \mathbb{R}^{\text{walk-length}}$ provide global structural context. The full node feature set is summarized in Table~\ref{tab:node_features}. The raw descriptors  $\mathbf{x}_i^0\in\mathbb{R}^{12+\text{walk-length}}$  are projected into a trainable latent space via a learned two layer MLP with layer normalization, $\text{LayerNorm}\left(\mathbf{\Theta}_2\sigma\left(\mathbf{\Theta}_1\mathbf{x}_i^0+\mathbf{b}_1\right)+\mathbf{b}_2\right)$, and the projected representation $\mathbf{x}_i^{\text{proj}}\in \mathbb{R}^{\text{proj}}$ is concatenated with the original signal: $\mathbf{x}_i^{\text{node}}=[\mathbf{x}_i^{\text{proj}} || \mathbf{x}_i^0]\in \mathbb{R}^{12+\text{walk-length+proj}}$.

\textbf{Edge feature enhancement}.
Edge features are initialized from the pairwise bonding matrix $M_{\text{init}}$ and enriched with a suite of physics-inspired and geometry-aware terms for the complete graph $G_{\text{complete}}$ encoding each message with both local metric strength and global environment context. The full edge feature set is given in Table~\ref{tab:edge_features}. We denote these as $\mathbf{e}_{ij}^0\in\mathbb{R}^{|\phi|+9}$ where $|\phi|$ is the number of radial basis functions (RBF) used.

\begin{table}
\renewcommand{\arraystretch}{1.3} 
\caption{Edge features using in both the GNN tracks}
\label{tab:edge_features}
\centering
\begin{tabularx}{\columnwidth}{l X} 
\toprule
\textbf{Feature} & \textbf{Description} \\
\midrule
$\phi(d_{ij})$ & Radial basis function (RBF) expansion of discrete distances for continuous behavior \\
$1/d_{ij}$ & Coulomb-like decay \\
$e^{-d_{ij}}$ & Exponential decay \\
$e^{-d_{ij}^2}$ & Gaussian decay \\
$d_{ij}$ & Raw distance \\
$\Phi_{ij}$ & Cosine similarity \\
$\mathrm{Var}[\Phi_{ij}]$ & Angular diversity \\
$M_{\text{init}}$ & Pairwise bonding according to $G$ \\
$|d_{\text{center},i} - d_{\text{center},j}|/d_{\max}$ & Atom positions relative to center \\
$(d_{\text{center},i} + d_{\text{center},j})/(2 d_{\max})$ & Average centrality of the pair \\
\bottomrule
\end{tabularx}
\end{table}

\subsubsection{Molecular Embedding} The enriched node representation $\mathbf{x}^{\text{node}}_i$ is processed by two independent GNN stacks running in parallel which we call \textbf{Dual-scale GNN} -- where each stack operates on a distinct neighborhood graph defined by its own cutoff radius $r_{\text{fine}}/r_{\text{coarse}}$. The fine-scale graph $G_{\text{fine}}$ retains atom pairs within $r_{\text{fine}}=2.5$ \AA \ capturing short-range, bond-level interactions while the coarse scale graph $G_{\text{coarse}}$ extends to $r_{\text{coarse}}=5$ \AA, encoding medium-range structural context. Each neighborhood graph is associated with its own edge features set $\mathbf{e}_{ij}^{\text{fine}}$ and $\mathbf{e}_{ij}^{\text{coarse}}$ respectively, constructed analogously to $\mathbf{e}_{ij}^0$ but for the two graphs $G_{\text{fine}}$ and $G_{\text{coarse}}$ respectively. The molecule $\tilde{\mathcal{M}}$ is thus mapped to two attributed graphs: $G_{\text{fine}}=\left(\mathcal{V}, \mathcal{E}_{\text{fine}}, \left\{ \mathbf{x}_i^{\text{node}} \right\}_{i=1}^N,\left\{\mathbf{e}_{ij}^{\text{fine}} \right\}_{(i, j) \in \mathcal{E}_{\text{fine}}}\right)$ and $G_{\text{coarse}}=\left(\mathcal{V}, \mathcal{E}_{\text{coarse}}, \left\{ \mathbf{x}_i^{\text{node}} \right\}_{i=1}^N,\left\{\mathbf{e}_{ij}^{\text{coarse}} \right\}_{(i, j) \in \mathcal{E}_{\text{coarse}}}\right)$.

Both graphs are independently processed by stacked GNN layers~\cite{gilmer2017neural, simonovsky2017dynamic} with residual connections, producing latent node embeddings for the $i^{\text{th}}$ node as $\mathbf{h}_i^{\text{fine}}, \mathbf{h}_i^{\text{coarse}} \in \mathbb{R}^{d_{\text{embed}}}$. The single layer node update rule for this GNN is 
\begin{equation}
    \mathbf{h}_i = \mathbf{\Theta_4} \mathbf{x}_i + \sum_{j \neq i} \text{MLP}_{\mathbf{\Theta_5}}(\mathbf{e}_{i, j}) \, \mathbf{x}_j,
\end{equation}
where $\mathbf{\Theta_4} \in \mathbb{R}^{d_\text{embed} \times d_x}$ is a learnable weight matrix and $\text{MLP}_{\mathbf{\Theta_5}} : \mathbb{R}^{d_e} \longmapsto \mathbb{R}^{d_\text{embed} \times d_x}$ generates edge-conditioned convolution kernels from the edge features. The resulting fine and coarse embeddings are concatenated and passed through a learnable fusion linear layer to produce a unified per-atom representation $\mathbf{x}^{\text{fuse}}_i\in\mathbb{R}^{\text{embed}}$:
\[
    \mathbf{x}^{\text{fuse}}_i = \sigma\left(\mathrm{LayerNorm}\left(\mathbf{\Theta}_3\left(\left[\mathbf{h}_i^{\text{fine}} \| \mathbf{h}_i^{\text{coarse}}\right]\right) + \mathbf{b}_3\right)\right).
\]

\begin{table}
\renewcommand{\arraystretch}{1.3} 
\caption{Pairwise geometry features for Readout MLP}
\label{tab:pairwise_mlp_features}
\centering
\begin{tabularx}{\columnwidth}{l X} 
\toprule
\textbf{Feature} & \textbf{Description} \\
\midrule
$|\mathbf{R}_{\text{PCA}(i)}-\mathbf{R}_{\text{PCA}(f)}|$ & Absolute difference in PCA projection of source $i$ and destination $j$ \\
$\Phi_{(c\to i)(i\to j)}$ & Cosine similarity between centroid $\to i$ and $i \to j$ \\
$\Phi_{(p\to i)(i\to j)}$ & Cosine triplet~\cite{mao2025molecule} between bond-partner according to G $\to i$ and $i \to j$ \\
$d_{ij}/d_{\max}$ & Fractional distance within the molecule \\
\bottomrule
\end{tabularx}
\end{table}

\subsubsection{Readout functions}

Generator predictions are created at the pair level from the fused node embeddings. For each ordered pair $(i < j)$ the fused node embeddings $\mathbf{x}_i^{\text{fuse}}$, $\mathbf{x}_j^{\text{fuse}}$ are combined with the initial edge features $\mathbf{e}_{ij}^0$ and a set of pairwise geometric descriptors  $\mathbf{e}_{ij}^{\text{pair}}\in\mathbb{R}^6$ (summarized in Table~\ref{tab:pairwise_mlp_features}).

The concatenated pair representation:
\[
    \left[\mathbf{x}_i^{\text{fuse}}+\mathbf{x}_j^{\text{fuse}}, \mathbf{x}_i^{\text{fuse}}-\mathbf{x}_j^{\text{fuse}}, \mathbf{x}_i^{\text{fuse}} \odot \mathbf{x}_j^{\text{fuse}}, \mathbf{e}_{ij}^0, \mathbf{e}_{ij}^{\text{pair}}\right] =: \mathbf{x}_{\text{pair}}
\]
is passed through a shared readout MLP $R : \mathbf{x}_{\text{pair}} \longmapsto \mathbb{R}$, which produces one scalar per pair, filling the entries of $\mathbf{A}_{\text{upper}}$:

\begin{equation}
    R(\mathbf{x}_{\text{pair}}) = \left\{(\mathbf{A})_{ij} = \text{MLP}_{\mathbf{\Theta_6}}(\mathbf{x}_{\text{pair}})\right\}_{i<j}.
\end{equation}

Since the readout MLP maps a fixed-dimensional pair representation to a scalar regardless of $N$, the model parameters are size agnostic and apply directly to molecules of any size. The full model is thus
\begin{equation}
\begin{split}
    F_{\mathbf{\Theta}, \mathbf{b}}: \mathcal{M} \xrightarrow[]{\text{pre-process}} \tilde{\mathcal{M}} \xrightarrow[]{\text{feat. eng.}} G_{\text{fine}}, G_{\text{coarse}} \\
    \xrightarrow[]{\text{GNN}} \hat{G}_{\text{fine}}, \hat{G}_{\text{coarse}} \xrightarrow[]{\text{MLP}} \mathbf{X}_{\text{pair}} \xrightarrow[]{R} \mathbf{A}_{\text{upper}} \xrightarrow[]{\exp(\cdot)} M_\text{oo}.
\end{split}
\end{equation}

\subsection{Training Method}

The model is trained on small hydrogenic systems -- $H_4$ and $H_6$ -- spanning a diverse set of molecular geometries, and evaluated on larger systems $H_8$, $H_{10}$ and $H_{12}$ without any retraining.

The design reflects a deliberate two-dimensional transferability objective. The primary goal is \emph{size transferability}: a model trained on a given geometry family at small system sizes should generalize to the same geometry family at larger sizes. A more ambitious secondary goal is cross-geometry transferability: a model trained on random geometries should generalize both across system sizes and across geometry families such as linear chains or ring configurations. Cross-geometry transfer proves to be a substantially harder problem for orbital coefficient prediction — in contrast to VQE circuit parameter prediction, where such generalization has already been demonstrated~\cite{bincoletto2025transferable} — and is left as a direction for future investigation. The experiments reported here focus on the size-transferability regime.

\textbf{Training objective}. The model is supervised using three complementary loss terms that together encourage recovery of both the correct occupied orbital subspace and the individual orbital structure, while remaining robust to the sign and permutation ambiguities intrinsic to orbital representations.

The primary loss is a \emph{Huber loss}~\cite{huber1992robust}, computed entry-wise between the predicted generator matrix $\mathbf{A}_{\text{upper}}$ and the reference generator $\mathbf{A}^{\text{ref}}_{\text{upper}}$. This loss combines advantages of both the $L_1$ (mean absolute error) and $L_2$ (mean squared error) loss, providing smoothness and gradient stability.

The two gauge-invariant losses complement this direct regression objective. The \emph{determinant overlap loss}  measures how well the occupied subspace spanned by the predicted orbitals aligns with that of the reference occupied subspaces. Given the occupied blocks $\hat{M}_{\text{occ}}$ and $M_{\text{occ}}^{\text{ref}}$
the overlap matrix $\mathbf{S} = \hat{M}_{\text{occ}}^\top M_\text{occ}^\text{ref}$
is formed and the loss is defined as $\mathcal{L}_\text{det} = 1 - |\det \mathbf{S}|^2$. This loss
vanishes when the two occupied subspaces coincide exactly and equals unity when they are orthogonal.
Crucially, this loss is invariant to unitary rotations within the occupied space, making it a
physically meaningful gauge-invariant measure of orbital quality.

An auxiliary \emph{sign-invariant orbital loss} $\mathcal{L}_\text{orb}$ complements this by
comparing individual orbital columns while accounting for the sign degree of freedom of each orbital. For each orbital, the loss takes the minimum squared Euclidean distance over the two
sign choices, and the result is averaged over all orbitals and batch samples. 
Together, these two losses encourage the model to recover both the correct occupied subspace
and the individual orbital structure, while remaining robust to the gauge degrees of freedom
inherent to the molecular orbital representation. The three terms are combined as
\begin{equation}
    \mathcal{L} = \mathcal{L}_{\text{Huber}} + \lambda_1\, \mathcal{L}_{\text{det}} + \lambda_2\, \mathcal{L}_{\text{orb}},
\end{equation}
where $\lambda_1$ and $\lambda_2$ are weighting hyper-parameters. The Huber loss directly supervises the generator entries, while the gauge-invariant terms steer the recovered orbital matrices towards physically meaningful solutions.

\section{Evaluation}
\label{sec:Eval}
We evaluate the proposed GNN framework along two axes: prediction accuracy on random molecular geometries, including transfer to unseen system sizes, and performance on structured geometric configurations where stronger electronic correlation presents a more challenging extrapolation regime. In both settings, we assess the quality of predicted orbital matrices both directly — by comparing predicted SPA energies against those obtained with classically optimized orbitals — and indirectly, by using the predicted matrices as warm-start initializations for a single classical optimization step.

\begin{figure}[p]
     \centering
     \begin{subfigure}[b]{\columnwidth}
         \centering
         \includegraphics[width=\textwidth]{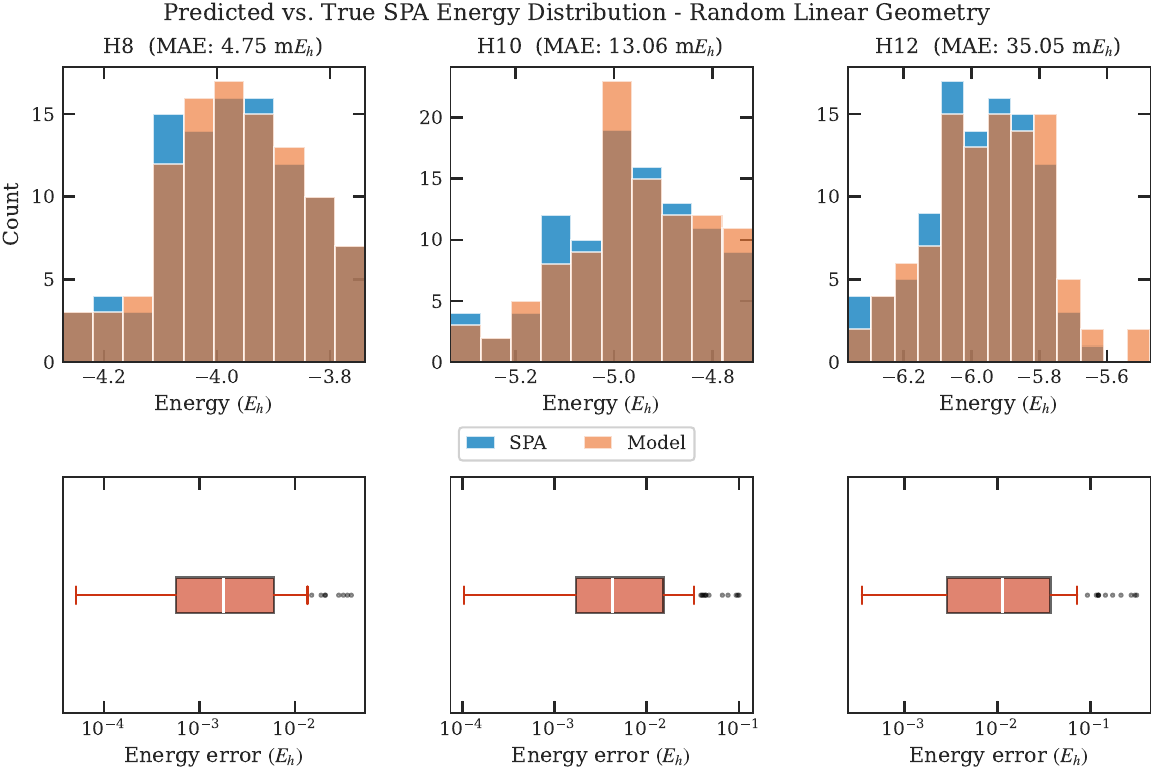}
         \caption{Random linear geometries}
     \end{subfigure}
     \hfill
     \begin{subfigure}[b]{\columnwidth}
         \centering
         \includegraphics[width=\textwidth]{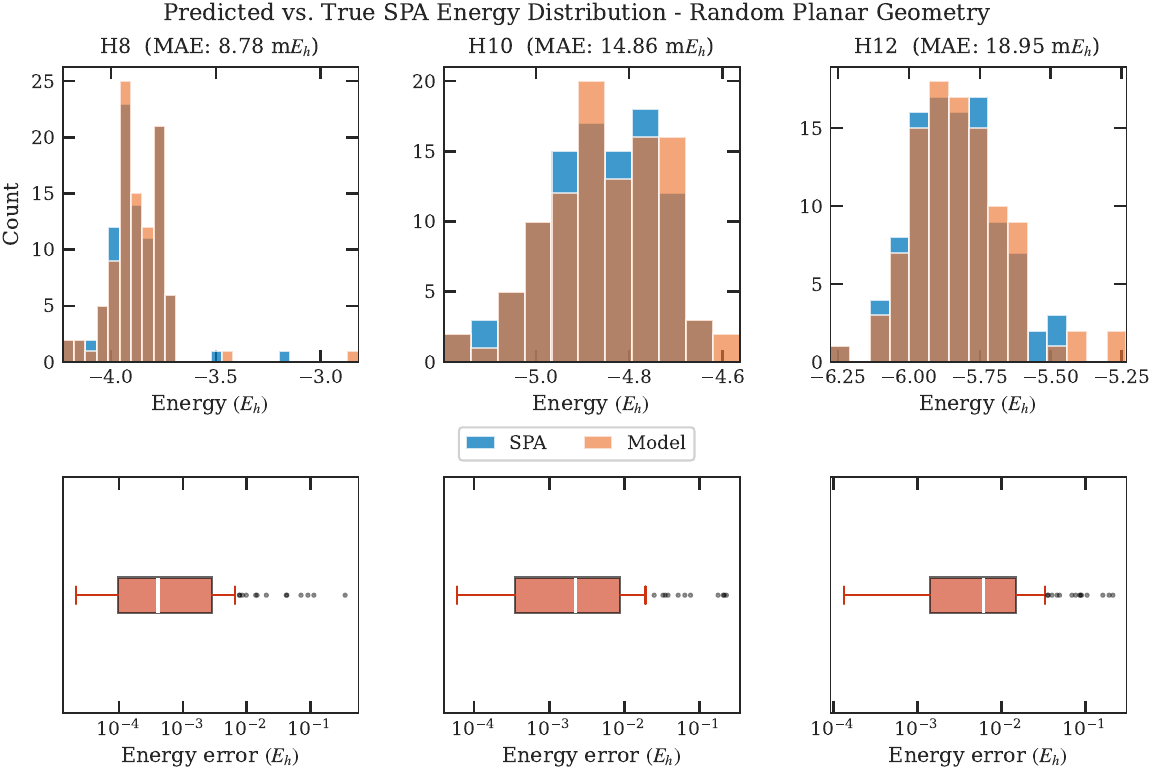}
         \caption{Random planar geometries}
     \end{subfigure}
     \hfill
     \begin{subfigure}[b]{\columnwidth}
         \centering
         \includegraphics[width=\textwidth]{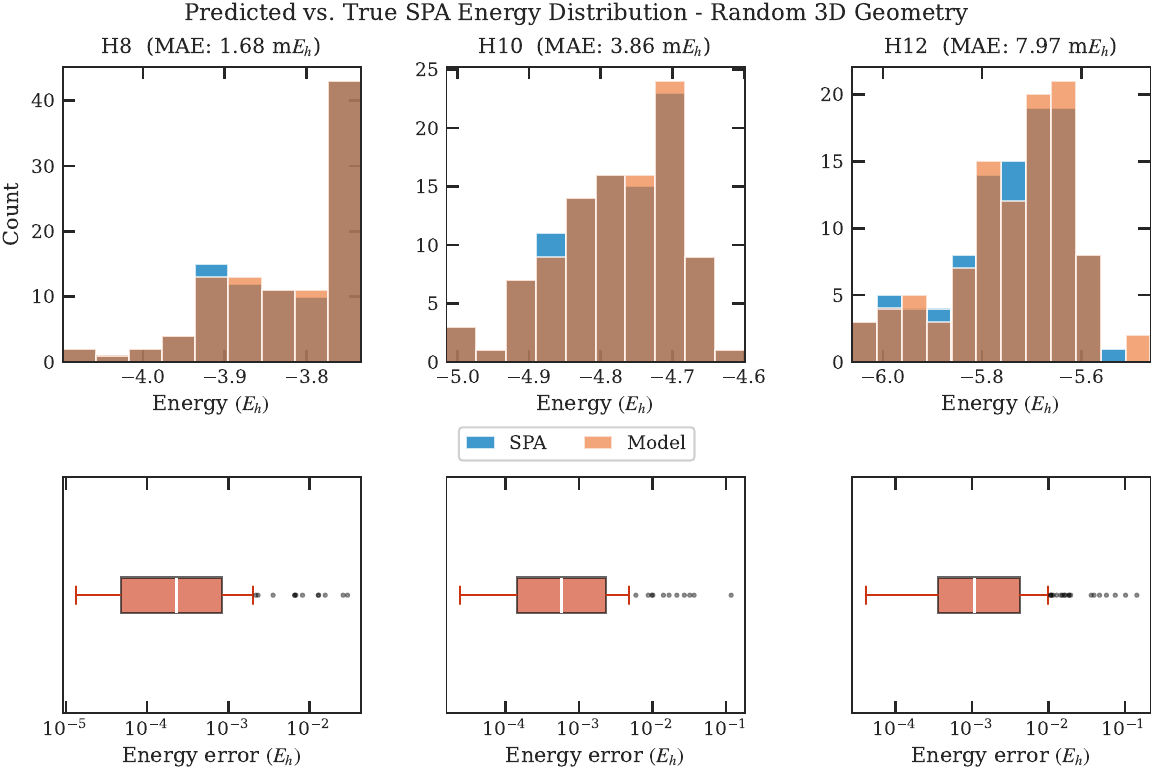}
         \caption{Random 3D geometries}
     \end{subfigure}
     \caption{\textbf{Predictions on random molecular geometries}. Energy distributions comparing SPA energies computed from predicted orbital matrices against the classically optimized reference for out-of-distribution system sizes, across three random geometry families: random linear (a), random planar (b), and random 3D (c). Our model achieves mean absolute energy errors of $\mathcal{O}(10) \ m E_h$ relative to the classically optimized reference, demonstrating that the architecture successfully generalizes the relationship between molecular geometry and optimal orbital structure to larger, unseen systems.}
     \label{fig: RandomGeoms}
\end{figure}

\begin{figure*}[!t]
     \centering
     \newcommand{\mysubsize}{\fontsize{8pt}{10pt}\selectfont}
     \begin{subfigure}[b]{0.48\textwidth}
         \centering
         \includegraphics[width=\textwidth]{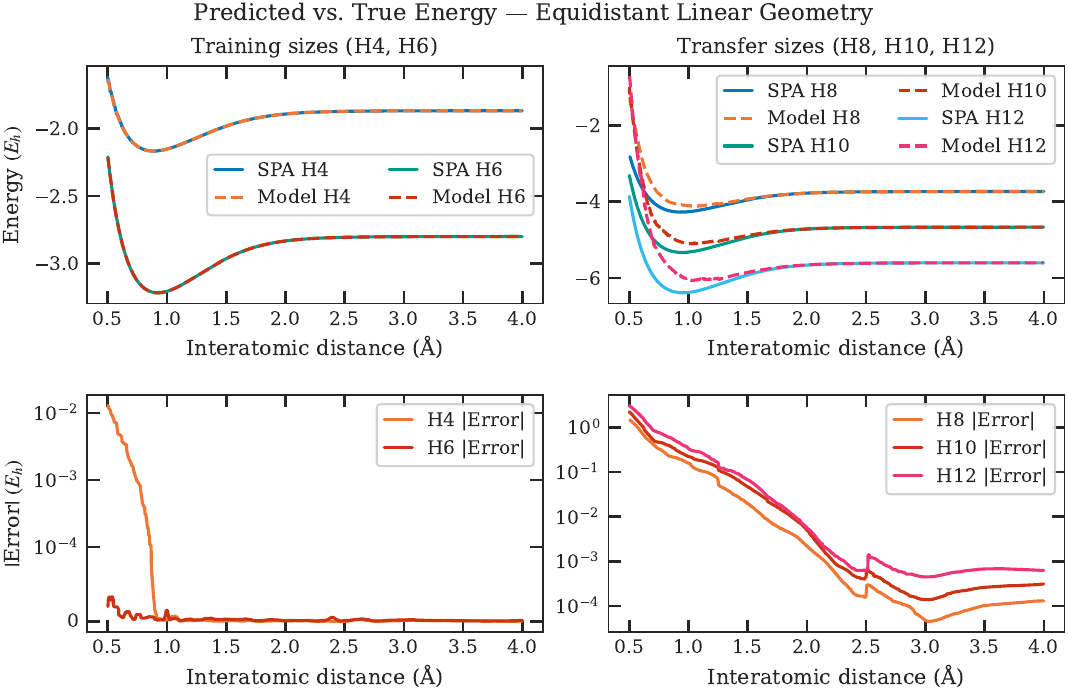}
         \subcaption{\mysubsize Direct prediction on Equidistant Linear Geometries}
     \end{subfigure}
     \hfill
     \begin{subfigure}[b]{0.48\textwidth}
         \centering
         \includegraphics[width=\textwidth]{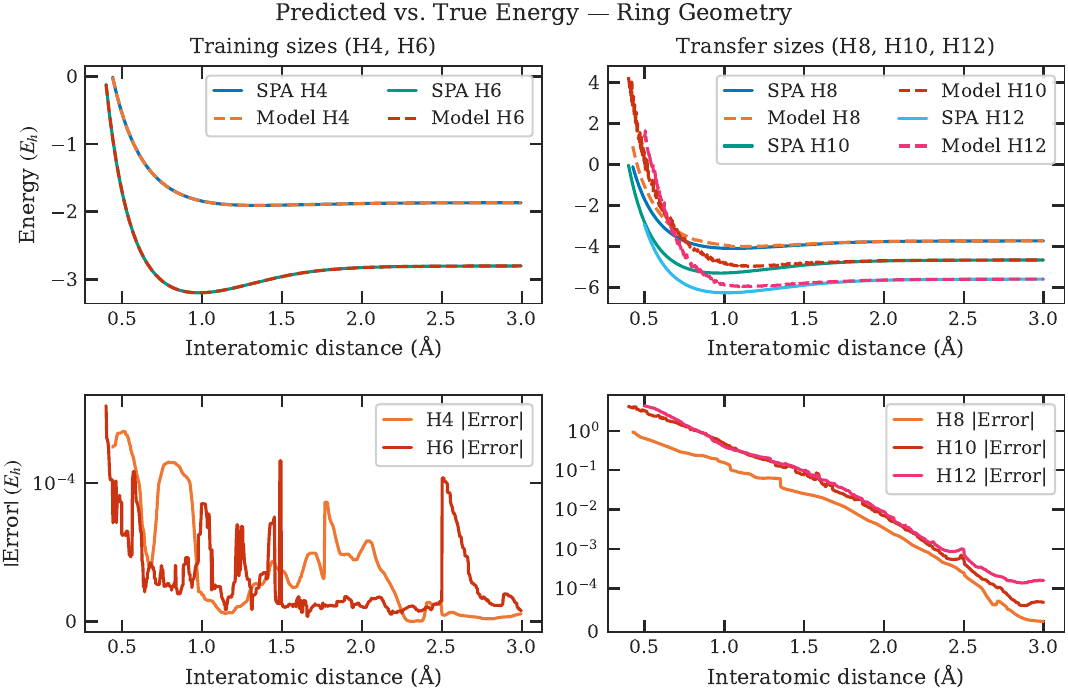}
         \subcaption{\mysubsize Direct prediction on Ring Geometries}
     \end{subfigure}
     \vspace{0.1cm} 
     \begin{subfigure}[b]{0.48\textwidth}
         \centering
         \includegraphics[width=\textwidth]{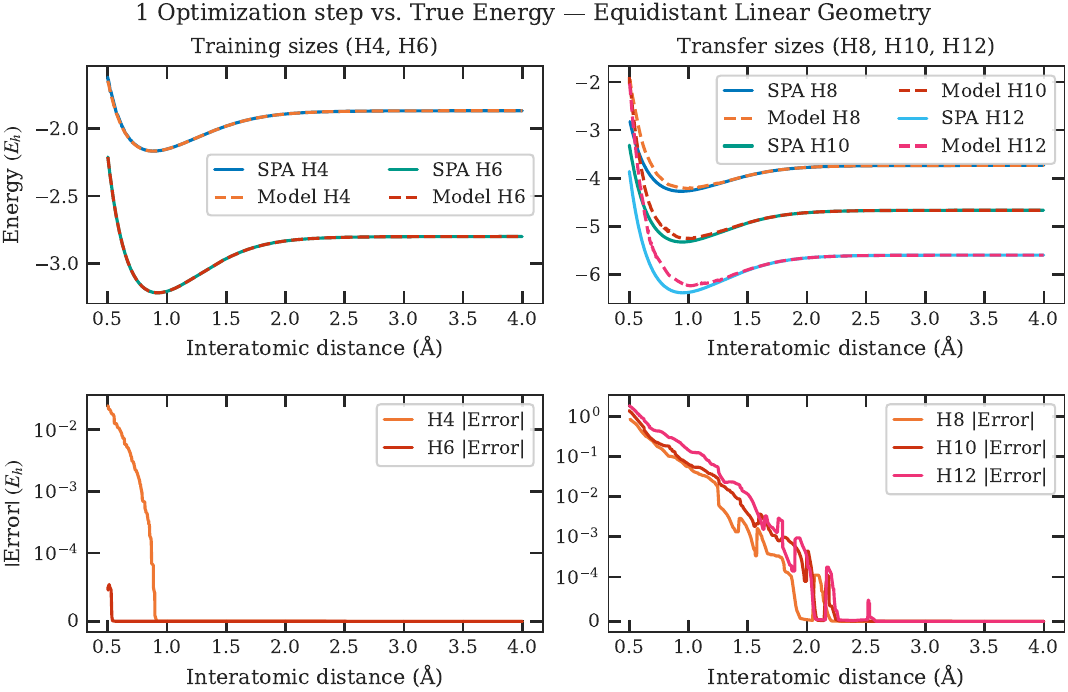}
         \subcaption{Single optimization step on Equidistant Linear geometries}
     \end{subfigure}
     \hfill
     \begin{subfigure}[b]{0.48\textwidth}
         \centering
         \includegraphics[width=\textwidth]{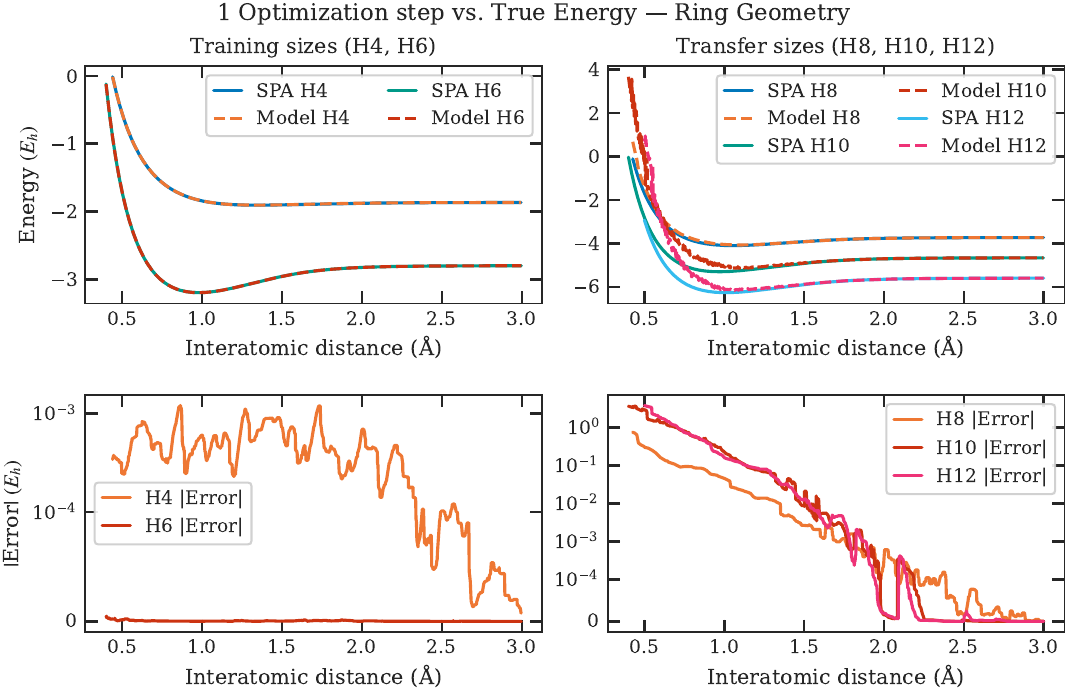}
         \subcaption{Single optimization step on Ring Geometries}
     \end{subfigure}
     \caption{\textbf{Predictions and warm-start evaluation on structured hydrogenic geometries}. Potential energy curves for equidistant linear configurations (a) and ring configurations (b), comparing SPA energies computed from predicted orbital matrices (dashed) against the classically optimized SPA reference (solid), across both in-distribution sizes ($H_4$ and $H_6$) and out-of-distribution sizes ($H_8$, $H_{10}$ and $H_{12}$). The model reproduces the reference curves closely in the dissociation regime $(1.5,4.0)$~\AA, while larger deviations are observed in the bonding regime $(0.5,1.5)$~\AA. For equidistant linear geometries, the mean absolute errors over the full dataset are $96.0$, $122.9$, and $170.0 \ mE_h$ for $H_8$, $H_{10}$ and $H_{12}$ respectively; for ring geometries, the corresponding errors are $119.7, 534.3$ and $461.8 \ mE_h$. Panels (c) and (d) show the effect of using model predictions as warm-start initializations for a single step of classical orbital optimization, for equidistant linear and ring geometries respectively. This hybrid strategy yields substantial error reductions across both geometry families: for equidistant linear geometries, the MAE decreases to $48.1, 60.1$, and $92.5 \ mE_h$ for $H_8$, $H_{10}$ and $H_{12}$ — corresponding to reductions of approximately 50\% relative to direct prediction. For ring geometries, the warm-start reduces errors to $50.3, 372.3$, and $268.8 \ mE_h$, with the most pronounced improvements observed for $H_8$, where the single optimization step recovers a large fraction of the remaining error. These results demonstrate that even moderately accurate orbital predictions provide a substantially better initialization than the default guess $M_{\text{init}}$, consistently reducing the number of optimizer iterations required to reach convergence.}
     \label{fig: StructredGeoms}
\end{figure*}

\subsection{Random Geometry Evaluation}
\label{sec:random_eval}
We first assess model performance on random molecular geometries across system sizes $N \in \{4,6,8,10,12\}$. The model is trained exclusively on $N \in \{4, 6 \}$; results on $N \in \{8,10,12\}$ therefore constitute a genuine out-of-distribution test of size transferability. We focus on three random geometry families -- random linear, random planar and random 3D -- as these represent the most physically relevant configurations. Results for $N \in \{8,10,12\}$ are shown in Fig.~\ref{fig: RandomGeoms}.

Performance is measured by the mean absolute error (MAE) between energies computed from model-predicted orbitals and the ground-truth SPA energies obtained via classical OO ($B$ is the batch of data  and $|B|$ is the batch size):
\begin{equation}
    \text{MAE} = \frac{\sum_{x\in B}|E^{\text{SPA}}_x - E^{\text{model}}_x|} {|B|}.
\end{equation}

Across all three random geometry families and both out-of-distribution system sizes, the model achieves energy accuracies of $\mathcal{O}(10) \ mE_h$, indicating that the predicted orbital matrices closely reproduce those obtained by classical optimization. These results confirm that the locally-defined features and size-agnostic readout described in Section~\ref{subsec:model_arch} successfully support generalization to larger systems. Notably, inference using the GNN model on average is $30\times$ faster in terms of the wall-clock time needed compared to  classical optimization routine per geometry instance, establishing it as a practical surrogate within the SPA pipeline.

\subsection{Structured Instance Evaluation}
\label{sec:struct_eval}
We next evaluate the model on two families of structured hydrogenic configurations: equidistant linear chains and ring geometries. For $H_8$, $H_{10}$ and $H_{12}$, configurations are generated with nearest neighbor spacings in $(0.5, 4.0)$ \AA, consistent with the training data distribution. Potential energy curves for all system sizes are shown in Fig.~\ref{fig: StructredGeoms}. On in-distribution sizes $H_4$ and $H_6$ the model reproduces the SPA potential energy curves accurately across the full range of interatomic distances, serving as a confirmation of the training procedure. The more informative test is on $H_8$, $H_{10}$ and $H_{12}$. In the dissociation regime $(1.5, 4.0)$ \AA, the model predicts orbital coefficients that yield SPA energies in close agreement with the classically optimized reference. However, in the bonding regime $(0.5, 1.5)$ \AA, the predicted energies deviate noticeably from the SPA ground truth. This degradation can be attributed to the strong orbital delocalization at short distance and degeneracy in the feature space due to the equidistant configuration, making it difficult for the model to capture the many-body effects — particularly for system sizes not seen during training.  

To mitigate this limitation, we explore using the model-predicted orbital matrix as a warm-start initialization for a single step of classical orbital optimization, rather than as a direct replacement. As shown in Fig.~\ref{fig: StructredGeoms}(c)-(d), this hybrid approach yields meaningful improvements in the bonding regime -- particularly for spacings in $(1.0, 1.5)$ \AA -- demonstrating that even imperfect orbital predictions provide a substantially better starting point than the default $M_{\text{init}}$. The warm-start strategy thus extends the practical utility of the model beyond the regime where direct prediction is accurate, reducing classical optimization iterations required to reach convergence.

Taken together, the results across both geometry families establish the proposed GNN framework as an effective and scalable approach to orbital prediction: delivering competitive surrogates for random geometries and large system sizes, and providing high-quality initializations that meaningfully reduce classical optimization cost for structured configurations.

\section{Conclusion \& Outlook}
\label{sec:Conc}
This work addresses a core bottleneck in the SPA-based VQE pipeline: the repeated, geometry-specific classical optimization of molecular orbital coefficients. We proposed a GNN framework that learns to predict optimized orbital coefficients directly from molecular geometry and pairwise bonding structure, bypassing explicit classical optimization at inference time. Trained on small hydrogenic systems ($H_4$ and $H_6$) across a geometrically diverse set of configurations, the model demonstrates strong size-transferability -- generalizing to larger, unseen systems ($H_8$, $H_{10}$ and $H_{12}$) without retraining. On random geometries, predicted orbital coefficients yield SPA energies within $\mathcal{O}(10) \ mE_h$ of the classically optimized reference across all tested system sizes, confirming that the locally-defined, size-agnostic architecture successfully captures the relationship between molecular geometry and electronic structure.

In the broader context of the SPA framework, this contribution represents the second step in a systematic effort to automate the full hybrid quantum-classical pipeline. The first step -- automating the initialization of SPA circuit parameters -- was established in the complementary work~\cite{bincoletto2025transferable}. Together, these two contributions replace the two most computationally demanding heuristic components of current VQE workflows with learned, transferable predictors: one targeting the orbital basis, the other targeting the variational circuit parameters. The result is a data-driven pipeline that substantially reduces the classical overhead of SPA-based VQE in particular, while providing a base for further developments targeting more general VQE approaches. Here we can imagine using the developed technology to also determine circuit parameters that can be interpreted as intermediate orbital rotations (see for example~\cite{kottmann2023molecular}).

\textbf{Limitations}. Despite these promising results, several limitations of our model exists. First, the model is trained and evaluated exclusively on minimal-basis hydrogenic systems using the STO-3G basis set, which limits the complexity of the electronic structure encountered during training. Instead of targeting larger basis sets individually, we suggest to investigate modern methods of direct determination with the machine learned orbitals in minimal basis as initial guesses~\cite{kottmann_direct_2020, valeev2023direct, langkabel2025advent}. Second, performance in the bonding regime remains a limitation for structured geometries, particularly for system sizes not seen during training. Third, the current model does not yet achieve cross-geometry transferability: a model trained on random geometries does not yet generalize reliably to structured geometric families such as equidistant linear chains or ring configurations, a capability that has been demonstrated for circuit parameter prediction but remains an open challenge for orbital prediction.

\textbf{Future directions}. Several natural extensions follow from this work. On the architectural side, incorporating higher-order geometric information — such as three-body angular terms or equivariant message passing — may improve performance in the bonding regime by better capturing non-local electronic interactions. Extending the framework to molecules beyond hydrogen, and to more accurate orbital bases, is a necessary step toward chemical relevance and is a direct priority for future work. This direction can be pursued following the framework of~\cite{kottmann2023molecular} which extends the graph-based SPA approach from hydrogenic systems to larger molecules by exploiting topological similarity across chemical species. Concretely, models trained on small hydrogen chains can be transferred to molecules sharing the same bonding topology: properties learned from $H_2$  inform predictions for $Li_2$ or the $\pi$-system of ethylene; $H_4$ transfers to $BeH_2$ or butadiene; and $H_6$ transfers to benzene. This topological transferability is not limited to orbital coefficient prediction but applies equally to circuit design and variational parameter prediction, suggesting a unified transfer learning strategy across the full SPA pipeline. The extended SPA framework~\cite{kottmann2024quantum} presents additional challenges with regards to automating heuristic components -- such as placement of the molecular gates when additional graphical structures are included to improve energy accuracy -- which represents a natural continuation of the automation pipeline. Finally, combining the orbital prediction model with the circuit parameter predictor of~\cite{bincoletto2025transferable} into a unified, end-to-end trainable system would close the remaining gap between data-driven initialization and full automation of the SPA-based VQE workflow.

\section*{Acknowledgment}
The author(s) used Claude for language editing in Sections ~\ref{sec:Methodology} and \ref{sec:Conc}. All content was edited and reviewed by the author(s), and take full responsibility of the final work. The authors also thank Francisco Javier Del Arco Santos for various fruitful discussions and Nico Meyer for valuable feedback on the manuscript.

\bibliographystyle{IEEEtran}
\bibliography{references}

\appendix
\begin{appendices}
\section{Separable Pair Approximation}
\label{sec:ApxSPA}
The Separable Pair Approximation (SPA)~\cite{kottmann2022optimized} is a variational circuit design that has demonstrated strong convergence and consistency for electronic structure problems. We restrict the following description to hydrogenic systems in minimal atomic bases and refer to the original works for the general molecular setting~\cite{kottmann2022optimized,kottmann2023molecular,kottmann2024quantum}. In SPA, the wave function is constructed as a tensor product of $N_e/2$ two-electron pair functions:
\begin{equation}\label{eq:SPA}
    |\Psi_{\text{SPA}}\rangle = \bigotimes_{k=1}^{N_e/2}|\psi_k\rangle.
\end{equation}
where each $|\psi_k\rangle$ is a linear combination of one-electron product states over a set of orbitals $S_k=\{\phi_l^k,l=0,...,|S_k|-1\}$
\begin{equation}
    \label{eq:pair_wfv}
    |\psi_k\rangle=\sum_{m,n\in |S_k|}c_{mn}^k|\phi_m^k\rangle\otimes|\phi_n^k\rangle.    
\end{equation}
This requires storing $\mathcal{O}(|S_k|^2)$ coefficients per pair. Applying the hard-core Boson (HCB) approximation encodes doubly occupied or unoccupied spatial orbitals into a single qubit reducing Eq.~\ref{eq:pair_wfv} to
$|\psi_k\rangle=\sum_{m\in |S_k|}c_{m}^k|\phi_{m\uparrow}^k\rangle\otimes|\phi_{m\downarrow}^k\rangle$,
lowering the memory requirement to $\mathcal{O}(|{S_k}|)$ and making the wavefunction classically simulable. A simple illustration is provided by $H_2$ in a minimal basis of two electrons in two orbitals. Starting from the reference $|1100\rangle$ a single two-electron excitation unitary yields the SPA wavefunction:
$\Psi_{\text{ref}} = |1100\rangle \xrightarrow{} \Psi_{\text{SPA}}= U_{\text{Exct}}\Psi_{\text{ref}}=e^{-i\frac{\theta}{2}Exct}|1100\rangle=c_0|1100\rangle+c_1|0011\rangle$,
where $\theta$ is the variational circuit parameter. The orbital sets $S_k$ are not uniquely determined a priori. One natural choice assigns each $S_k$ as a canonical Hartree-Fock orbital paired with its anti-bonding counterpart, recovering part of the static correlation~\cite{henderson2014seniority}. A more systematic approach draws on the molecular graph (Lewis structure) concept from Valence Bond theory~\cite{2007Mapping}; candidate graphs are constructed following Rumer's rule, the most relevant is selected to define the ansatz, and the orbitals are subsequently optimized to minimize the SPA energy.
Finally, SPA is not an isolated ansatz but appears as a substructure in a range of VQE approaches, including graph-based circuit designs~\cite{kottmann2023molecular}, tiled-unitary products~\cite{burton2024accurate}, quantum number preserving fabrics~\cite{anselmetti2021local} and decomposed linear combinations of SPA circuits~\cite{kottmann2024quantum}.

\section{Orbital Optimization}
\label{sec:ApxOrbCoeff}
\begin{figure}[htbp]
    \centering
    
    \begin{subfigure}[b]{0.45\textwidth}
        \centering
        $
        \begin{bmatrix}
          1.027 & -0.138 &  0.018 & -0.002 \\
         -0.138 &  1.055 & -0.142 &  0.018 \\
          0.018 & -0.142 &  1.055 & -0.138 \\
         -0.002 &  0.018 & -0.138 &  1.027 
        \end{bmatrix}
        $
        
        $\downarrow$

        \begin{tabular}{cccc}
            \rotatebox{-90}{\includegraphics[width=0.3\textwidth]{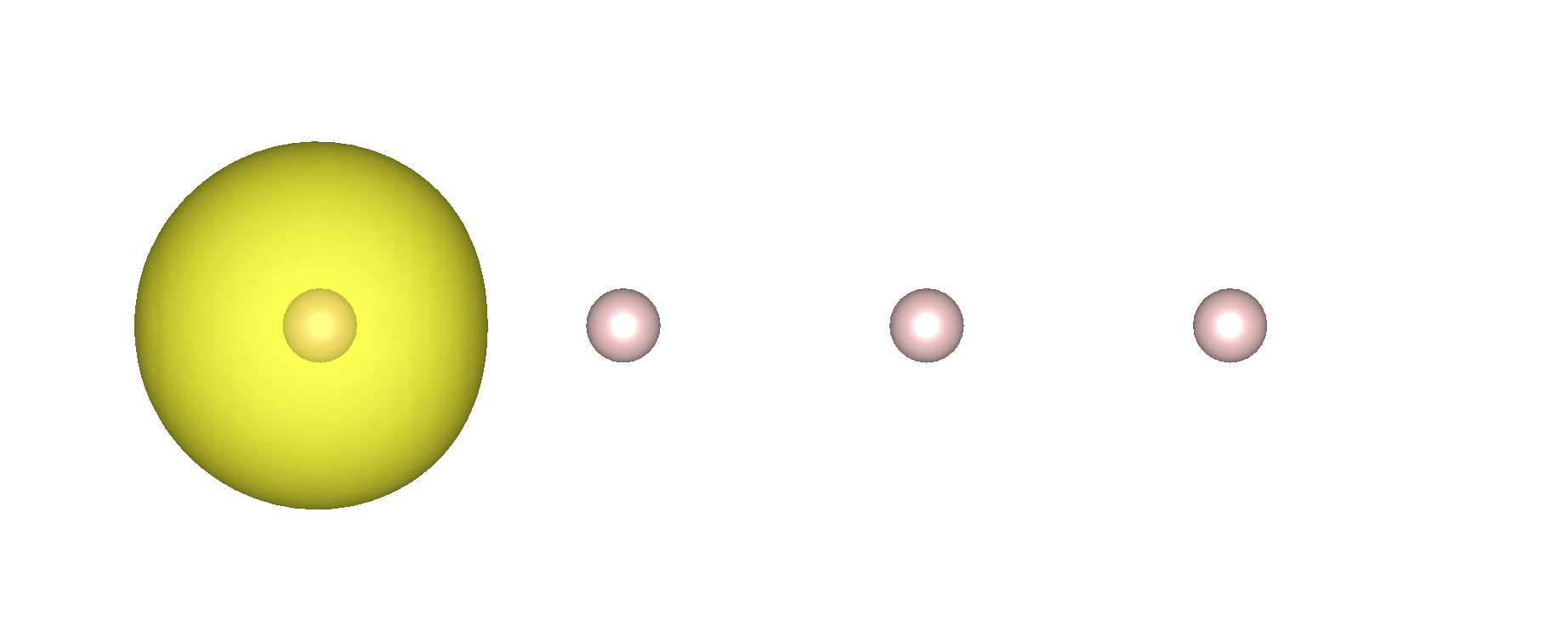}} &
            \rotatebox{-90}{\includegraphics[width=0.3\textwidth]{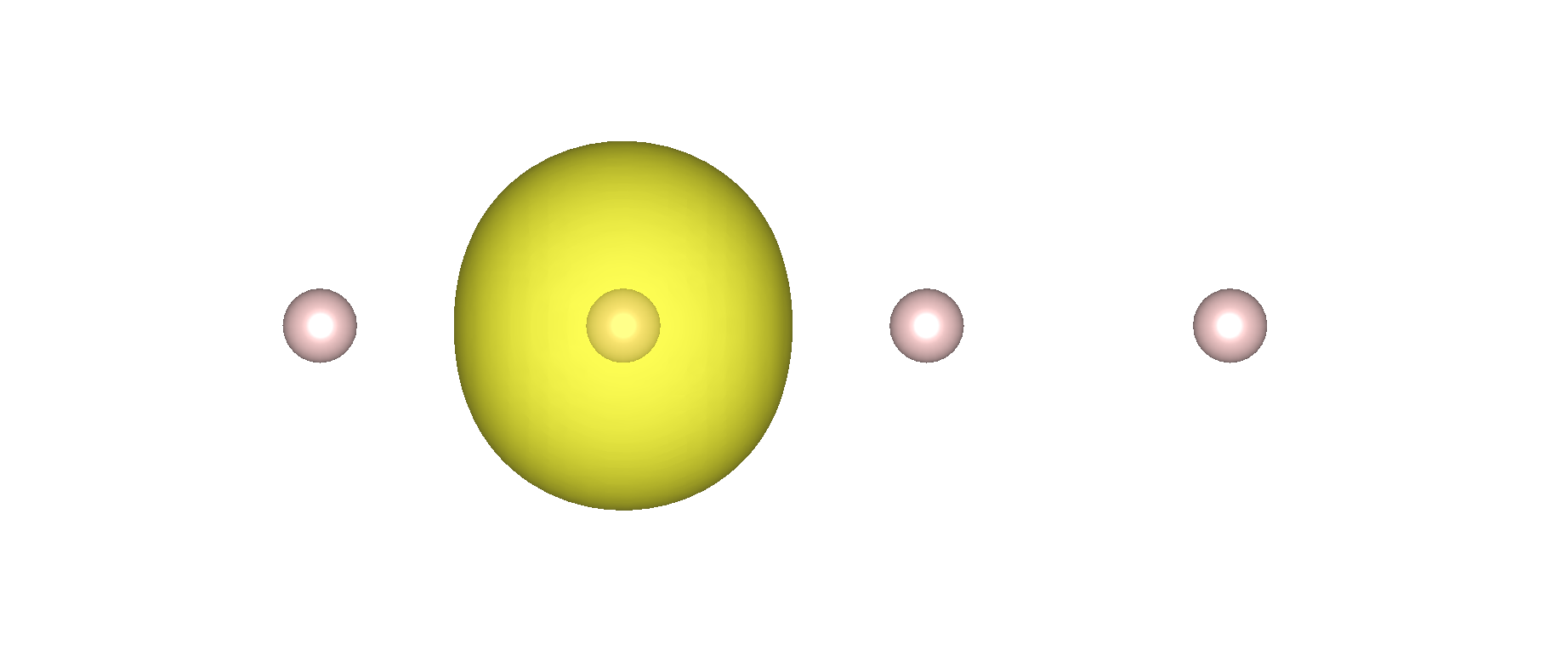}} &
            \rotatebox{-90}{\includegraphics[width=0.3\textwidth]{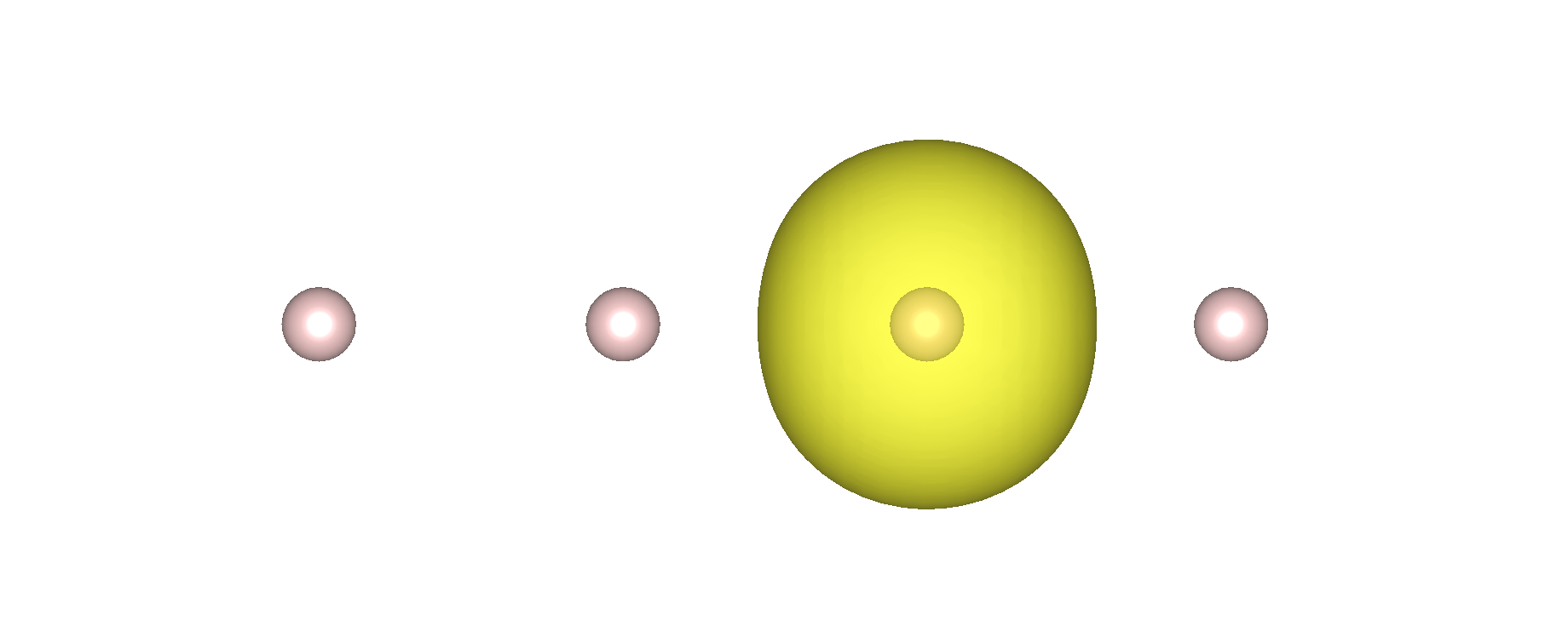}} &
            \rotatebox{-90}{\includegraphics[width=0.3\textwidth]{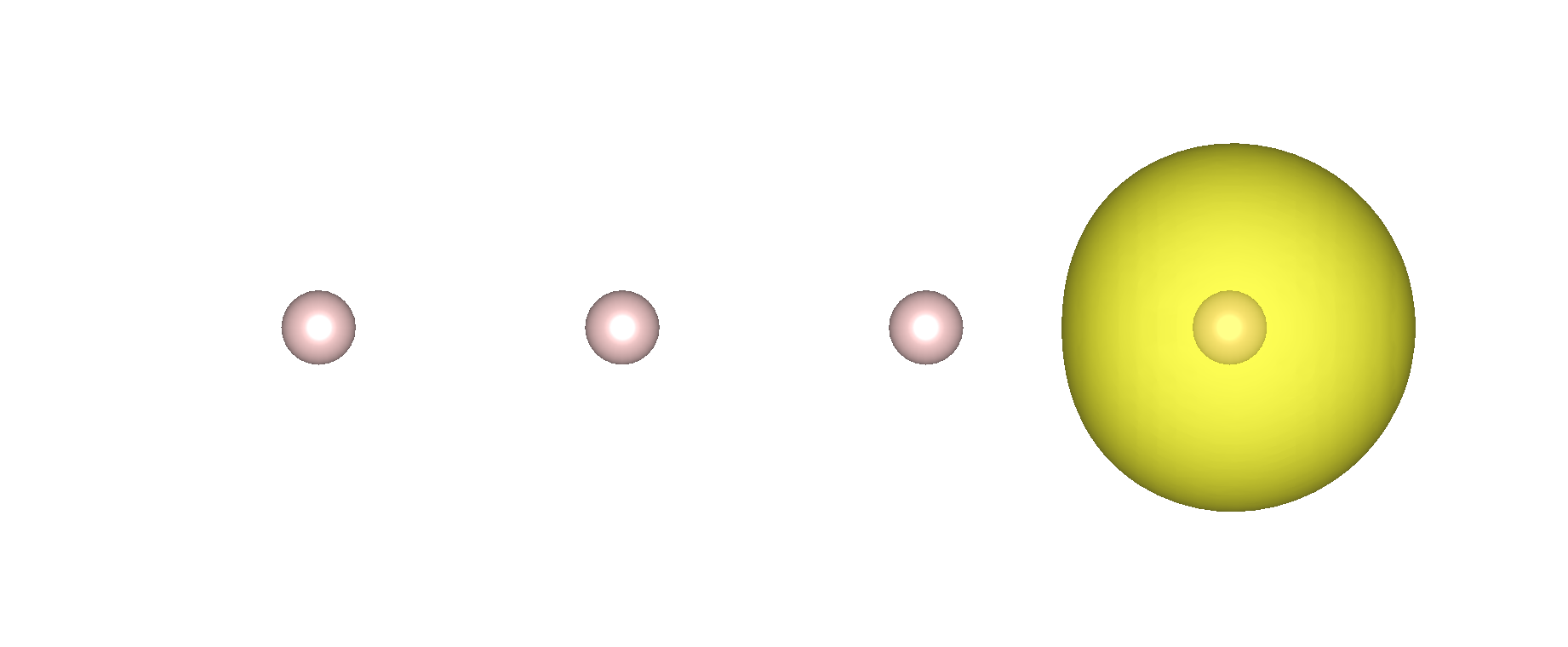}}
        \end{tabular}
        
    \end{subfigure}
    \hfill
    \begin{subfigure}[b]{0.45\textwidth}
        \centering
        $
        \begin{bmatrix}
          0.818 &  0.628 & -0.059 & -0.085 \\
         -0.850 &  0.628 & -0.188 &  0.011 \\
          0.188 &  0.011 &  0.850 &  0.628 \\
          0.059 & -0.085 & -0.818 &  0.628 
        \end{bmatrix}
        $
        
        $\downarrow$
        
        \begin{tabular}{cccc}
            \rotatebox{-90}{\includegraphics[width=0.3\textwidth]{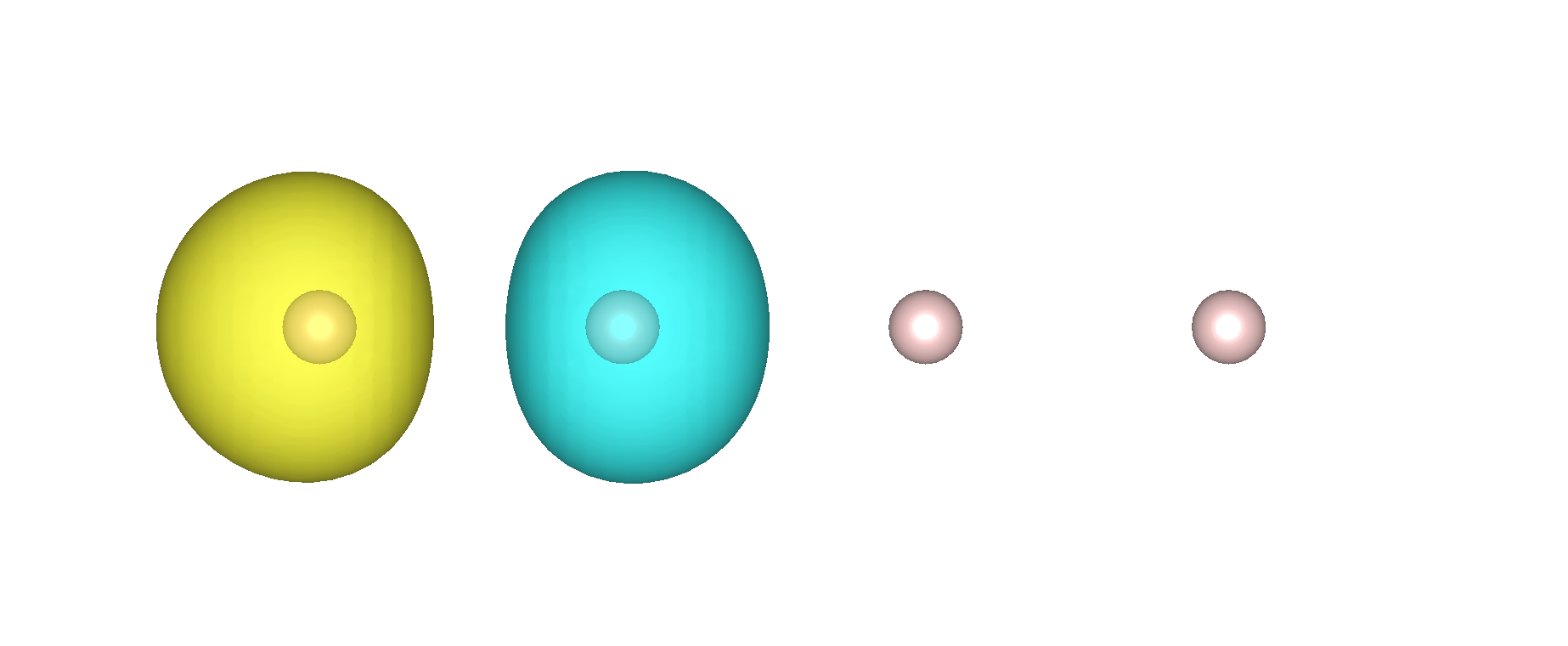}} &
            \rotatebox{-90}{\includegraphics[width=0.3\textwidth]{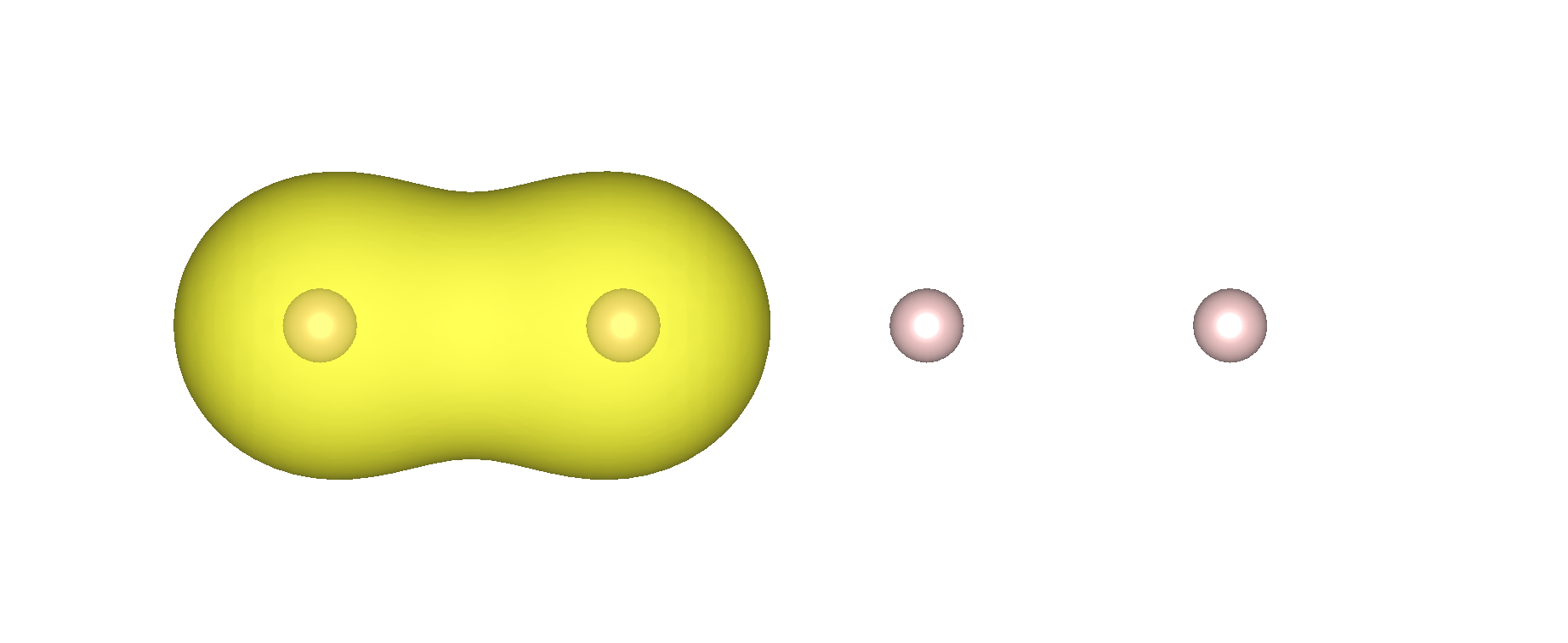}} &
            \rotatebox{-90}{\includegraphics[width=0.3\textwidth]{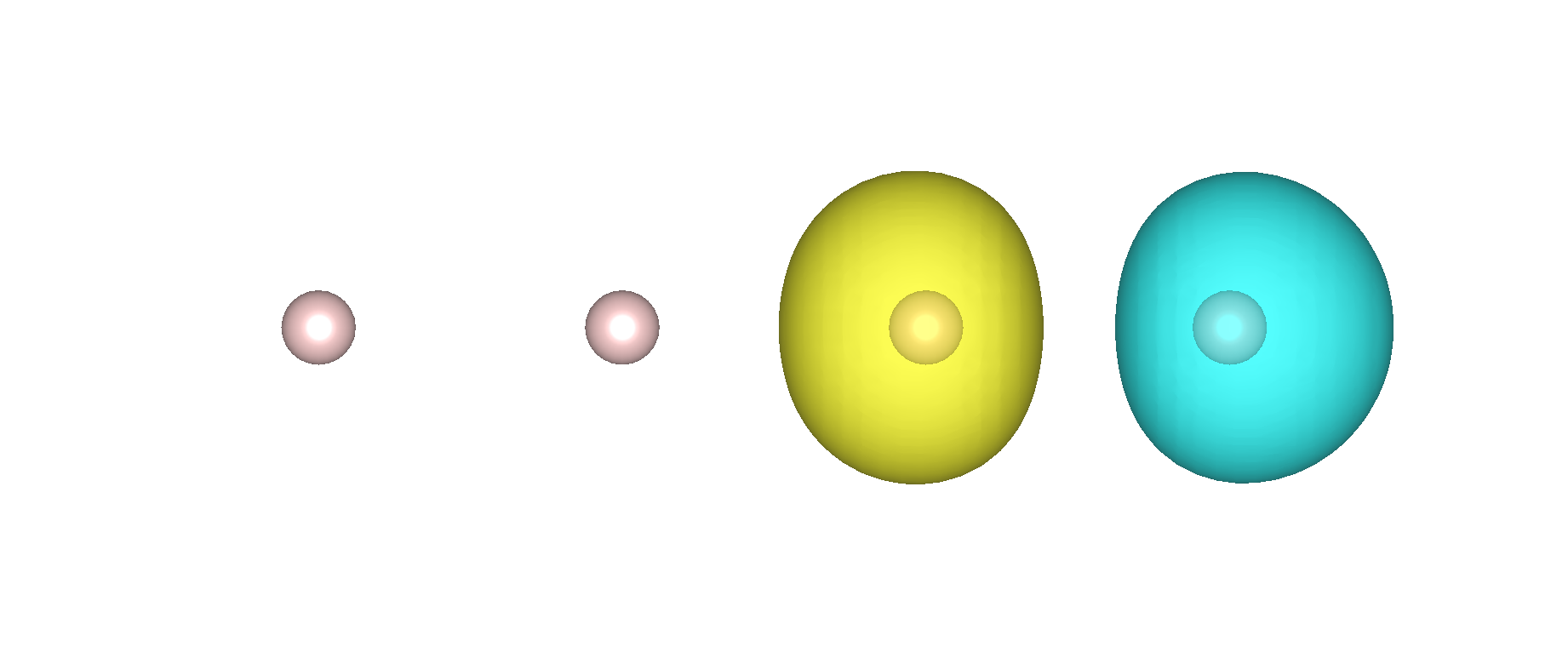}} &
            \rotatebox{-90}{\includegraphics[width=0.3\textwidth]{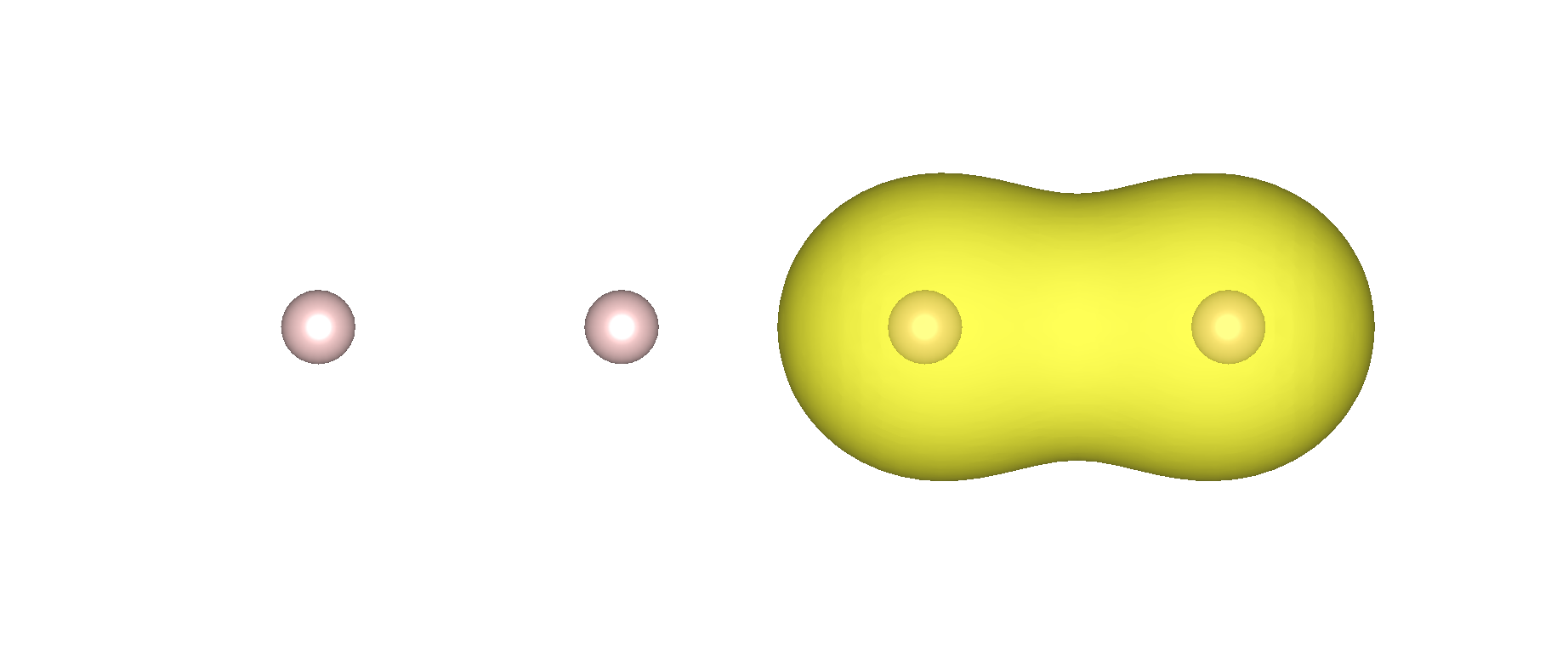}}
        \end{tabular}
        
    \end{subfigure}
    \caption{Visualization of Native (top) and SPA (bottom) orbitals of the $H_4$ molecule. Each 3D visualization represents one column of the respective orbital coefficient matrix.}
    \label{fig:orbital_visualization}
\end{figure}

Orbital optimization is the process of finding the optimal linear combinations of atomic basis functions for a given ansatz. Mathematically, this corresponds to applying a unitary rotation to the initial orbital set, mixing the basis functions into new combinations. Each such rotation is parameterized by a set of angles, and the optimization objective is to find the angle values that minimize the total energy of the ansatz~\cite{mizukami2020orbital}. 
The key consequence of an orbital rotation is that it transforms the one- and two-electron integrals encoding the kinetic energy, nuclear attraction, and electron-electron repulsion of the system — effectively producing a molecule-specific Hamiltonian maximally adapted to the ansatz. Crucially, this transformation is performed entirely classically: the rotation is pre-applied to the integral tensors before circuit execution, absorbing part of the unitary into the Hamiltonian rather than implementing it as additional quantum gates. This keeps the quantum circuit shallower, ensuring the quantum processor executes only the correlated operations that cannot be replicated classically. Fig.~\ref{fig:orbital_visualization} illustrates the effect by showing the native and SPA-optimized orbitals of the $H_4$ molecule.
\section{Further Feature Engineering Details}
\textbf{Centroid}. The atomic coordinates $\mathbf{R}=\{\mathbf{r}_i \}_{i=1}^N$ are translated to the molecular centroid, 
\[
\overline{\mathbf{c}} = \frac{1}{N} \sum^N_{i=1}\mathbf{r}_i, \quad \mathbf{r}_i \leftarrow \mathbf{r}_i - \overline{\mathbf{c}}
\]
ensuring that all positional features are defined relative to the molecular center of mass.

\textbf{Angular Information}.
To capture the directional geometry of the molecular environment, we define the cosine similarity matrix $\Phi \in [-1, 1]^{N \times N}$ with entries
\[
\Phi_{(i\to\bar{\mathbf{c}})(j\to\bar{\mathbf{c}})} = \frac{\mathbf{r}_{i\bar{\mathbf{c}}}^\top \mathbf{r}_{j\bar{\mathbf{c}}}}{||\mathbf{r}_{i\bar{\mathbf{c}}}||_2 \, ||\mathbf{r}_{j\bar{\mathbf{c}}}||_2} = \cos{(\phi_{ij})},
\]
where $\phi_{ij}$ is the angle subtended between atoms $i$ and $j$ relative to the molecular centroid. For each atom $i$, we additionally compute the variance of its row in $\Phi$, which we refer to as the \emph{angular diversity}. This scalar quantifies how isotropic the local atomic environment is: a low variance indicates a quasi-linear arrangement of neighbors, while a high variance is characteristic of a 3D environment.

\textbf{PCA-Based Positional Encoding}: Each atom is assigned a size-invariant position relative to its molecular frame by projecting the centroid-centered coordinates onto the three principal axes of the molecule, computed per graph via singular value decomposition (SVD). The resulting projections are normalized to $[0,1]$, yielding a three-dimensional encoding that captures each atom's position along the dominant structural directions of the molecule without depending on absolute coordinates or system size.

\textbf{Random Walk Structural Encoding (RWSE)}. Structural position within the molecular graph is encoded using the diagonal entries of successive powers of the random-walk transition matrix
$P = AD^{-1}$ (where $A$ is the adjacency matrix of the radius-graph $G_{\text{cut}}$ and $D$ is the  corresponding degree matrix). The encoding for node $i$ is
\[
p_i^{\text{RWSE}} = \bigl[(P^1)_{ii},\; (P^2)_{ii},\; \ldots,\; (P^T)_{ii}\bigr]\in \mathbb{R}^T,
\]
where $(P^k)_{ii}$ is the probability of a random walk returning to node $i$ after exactly $k$ steps. Short walk lengths capture local connectivity structure, while longer walks encode progressively more global topological context — providing a multi-scale structural fingerprint that is defined purely from graph topology and is therefore size-agnostic.

\textbf{Radial Basis Function (RBF) Expansion}: Scalar pairwise distances are encoded into smooth, localized continuous representations using a set of $l$ Gaussian basis functions uniformly spaced over $[d_{\min}, d_{\max}]$: 
\[
\phi_k(r) = \exp\!\left(-\frac{(r - \mu_k)^2}{2\sigma^2}\right),
    \quad k = 1,\ldots,l,
\]
with width $\sigma = (d_{\max}-d_{\min})/l$. Replacing raw scalar distances with an RBF expansion ensures that distance information is encoded continuously and smoothly.

\begin{table}[htbp]
\caption{Summary of training hyperparameters.}
\begin{center}
\begin{tabular}{|c|c|c|c|}
\hline
\textbf{Hyperparameter}&\multicolumn{3}{|c|}{\textbf{Value}} \\
\hline
\cline{2-4} 
Optimiser                          & Adam \\
Initial learning rate $\eta_0$     & $10^{-4}$ \\
Training epochs                    & 1000 \\
\midrule
GNN message-passing layers         & 3 \\
GNN hidden dimension               & 64 \\
Output MLP layers                  & 2 \\
Output MLP hidden dimension        & 128 \\
RBF basis functions                & 20 \\
Interaction cutoff $r_\mathrm{\text{fine}}$& 2.5\,\AA \\
Interaction cutoff $r_\mathrm{\text{coarse}}$& 5.0\,\AA \\
\midrule
Training dataset size              & 224k \\
\hline
\end{tabular}
\label{tab:hyperparams}
\end{center}
\end{table}

\clearpage
\onecolumn
\section*{Code Implementation}
\begin{lstlisting}[language=Python, caption={Tequila H4 example}]
import tequila as tq
import numpy as np

geometry = """
H 0.0 0.0 1.5
H 0.0 0.0 3.0
H 0.0 0.0 4.5
H 0.0 0.0 6.0
"""

graph = [(0, 1), (2, 3)]

def energy_calculator(geometry: str, graph: List[Tuple[int, int]]):

    mol = tq.Molecule(geometry = geometry, basis_set="sto-3g")
    U = mol.make_ansatz(name="HCB-SPA", edges=edges)
    
    initial_guess = np.eye(4)
    for edge in edges:
        initial_guess[edge[0]][edge[1]] = 1.0
        initial_guess[edge[1]][edge[0]] = -1.0
    
    opt_orbitals = tq.chemistry.optimize_orbitals(
                        molecule.use_native_orbitals(),
                        U,
                        initial_guess=initial_guess,
                        silent=True,
                        use_hcb=True,
                   )
    
    M_oo = opt_orbitals.mo_coeff
    
    mol = opt.molecule
    H = mol.make_hardcore_boson_hamiltonian()
    E = tq.ExpectationValue(H=H, U=U)
    energy = tq.minimize(E, silent=True)

    return energy, M_oo 
    
\end{lstlisting}

\end{appendices}

\end{document}